\documentstyle[psfig,epsf]{mn}

\begin{document}

\title[Peculiar velocities of galaxies and clusters]
{Peculiar velocities of galaxies and clusters}
\author[Ravi K. Sheth \& Antonaldo Diaferio]
{Ravi K. Sheth$^1$ \& Antonaldo Diaferio$^2$\\
$^1$ NASA/Fermilab Astrophysics Group, Batavia, IL 60510-0500\\
$^2$ Dipartimento di Fisica Generale ``Amedeo Avogadro'', 
Universit\`a di Torino, Italy\\
\smallskip
Email: sheth@fnal.gov, diaferio@ph.unito.it
}
\date{Submitted to MNRAS 2000 July 15}

\maketitle

\begin{abstract}
We present a simple model for the shape of the distribution
function of galaxy peculiar velocities.  We show how both nonlinear 
and linear theory terms combine to produce a distribution which has an 
approximately Gaussian core with exponential wings.  The model is easily 
extended to study how the statistic depends on the type of particle used 
to trace the velocity field (dark matter particles, dark matter haloes, 
galaxies), and on the density of the environment in which the test 
particles are.  Comparisons with simulations suggest that our model 
is accurate.  We also show that the evolution of the peculiar velocities 
depends on the local, rather than the global density.  Since clusters
populate denser regions on average, using cluster velocities with the 
linear theory scaling may lead to an overestimate of the global value 
of $\Omega_0$. Conversely, using linear theory with the global value of 
$\Omega_0$ to scale cluster velocities from the initial to the present 
time results in an underestimate of their true velocities.  In general, 
however, the directions of motions of haloes are rather well described 
by linear theory.  Our results help to simplify models of redshift-space 
distortions considerably.  
\end{abstract}

\begin{keywords}  galaxies: clustering -- cosmology: theory -- dark matter.
\end{keywords}

\section{Introduction}\label{intro}
The gravitational evolution of the density field in an expanding 
Universe modifies the shape of the distribution function $f(v)\,{\rm d}v$ 
of the peculiar velocity field.  Because of the action of 
gravity, the dark matter distribution at the present time is 
certainly not an ideal gas, so it should come as no surprise that the 
present-day distribution of velocities is quite different from the 
Maxwell-Boltzmann.  Although it has been known for some time that the 
variance of this distribution is sensitive to the background cosmology, 
the first detailed model of how and why the shape of this distribution 
differs from a Maxwell-Boltzmann was discussed by Saslaw et al. (1990).  

Here we present a model for the shape of the peculiar velocity 
distribution $f(v)\,{\rm d}v$ which captures the essence of the evolution 
driven by gravitational instability. Our model explicitly includes the 
type of particle tracing the velocity field: dark matter particles, 
galaxies, or galaxy clusters. This feature of the model is crucial 
because galaxies and clusters trace the underlying distribution of 
dark matter differently and measurements of $f(v)\,{\rm d}v$, in which 
either galaxies or galaxy clusters are used as trace particles 
(e.g., the Mark III sample of Willick et al. 1997, 
the SFI sample of Giovanelli et al. 1998 and Haynes et al. 1999a,b, 
and the ENEAR sample of da Costa et al. 2000)
are widely used to constrain cosmological models.

Section~\ref{model} describes our model.  In essence, we write the motion 
of a dark matter particle as the sum of two terms, one of which evolves 
according to linear theory, and another which is inherently nonlinear.  
Section~\ref{virhalo} shows that this split provides a good approximation 
to what happens in N-body simulations, and Section~\ref{fvdm} shows that 
our model provides a good description of the shape of $f(v)$ for the 
dark matter.  Section~\ref{fvden} describes how the model can be used 
to study the dependence of the shape of $f(v)$ on the local density, and 
Section~\ref{fvgal} shows how and why the distribution of halo velocities 
is different from that of dark matter particles.  This leads, in 
Section~\ref{fvgal}, to a discussion of how the model can be extended to 
provide predictions for how $f(v)$ depends on galaxy type.  

In addition to studying velocities at the present time, we also study 
how velocities evolve from some early time to the present.  
Section~\ref{evolve} argues that the evolution of halo speeds 
should show some dependence on local density, and presents results 
from simulations showing that this actually does happen.  We also study 
how the direction of motion of trace particles evolves.  
We show that linear theory describes the directions in which dark matter 
particles are moving today rather badly, but the directions of halo motions 
very well, and discuss some useful consequences of this fact.  
We summarize our findings in Section~\ref{discuss}.

\section{The distribution function of velocities}\label{model}
The key assumptions of this model are similar to those discussed by 
Sheth (1996) and Diaferio \& Geller (1996) in their discussions of 
the distribution of pairwise velocities.  Indeed, the model described 
below is essentially the same as that discussed in the second half of 
Sheth (1996), where a detailed discussion of the significant differences 
between our approach and that of Saslaw et al. (1990) can also be found.  
The final subsection discusses how the model can be extended to describe 
what happens when dark matter haloes or galaxies, rather than dark 
matter particles, are used to construct the statistic.  

\subsection{Assumptions}\label{virhalo}
All dark matter particles are assumed to be in approximately spherical, 
virialized haloes.  The velocity of a dark matter particle is assumed 
to depend on two variables:  the mass of the halo in which it is, and 
the local density in which the parent halo is.  Although the local 
density is, in principle, a function of smoothing scale, we show below 
that if this scale is chosen so that it typically contains many halos, 
then our results are relatively insensitive to the exact choice.  
Let $p(v|m,\delta)\,{\rm d}v$ denote the probability that a particle 
in a halo of mass $m$ which is in a region within which the average 
density is $(1+\delta)$ times the background density, has velocity in 
the range d$v$ about $v$.  
Then the distribution of interest is given by summing up the various 
$p(v|m,\delta)$ distributions, weighting by the fraction of particles 
which are in haloes of mass $m$ in regions of overdensity $\delta$:  
\begin{equation}
 f(v) = {\int {\rm d}\delta\,p(\delta)
         \int {\rm d}m\,m\,n(m|\delta)\,p(v|m,\delta)
 \over   \int {\rm d}\delta\,p(\delta)\int {\rm d}m\,m\,n(m|\delta)},
\label{fvexact}
\end{equation}
where $n(m|\delta)\,{\rm d}m$ is the number density of haloes that 
have mass in the range d$m$ about $m$ and are in regions within which 
the average density is $\delta$, and $p(\delta)\,{\rm d}\delta$ is 
the fraction of regions which have density in the range d$\delta$ 
about $\delta$.  The weighting by $m$ reflects the fact that the 
number of dark matter particles in a halo is supposed to be proportional 
to the halo mass.  This expression holds both for the size of the 
velocity vector itself, which we will often call the speed, as well as 
for the individual velocity components.  

If we rearrange the order of the integrals in the denominator, and then 
integrate over $\delta$, then what remains is $\int {\rm d}m\,m\,n(m)$, 
where $n(m)$ is the average density of haloes of mass $m$ averaged over 
all environments, so it is often called the universal halo mass function.  
Because all particles are assumed to be in haloes, this universal mass 
function is normalized so that $\int {\rm d}m\,m\,n(m)$ equals the 
background density $\bar\rho$.  

To proceed, we need estimates of $n(m|\delta)$ and of $p(v|m,\delta)$.  
In what follows, we will study the case of clustering from Gaussian 
initial conditions in detail. If the initial conditions are 
non-Gaussian, $n(m|\delta)$ and $p(v|m,\delta)$ are of course different 
from what we use, but our equation~(\ref{fvexact}) is still valid.  

For hierarchical clustering from Gaussian initial conditions, a good a
pproximation to the universal number density $n(m)$ of haloes of mass 
$m$ is given, for arbitrary power-spectra, cosmology and times, by 
Press \& Schechter (1974).  In what follows we will use the simple 
modification to this formula provided by Sheth \& Tormen (1999) which 
is considerably more accurate:
\begin{equation}
 n(m) = {A\bar\rho\over m^2}\,\left(1 + {1\over (a\nu^2)^p}\right) 
  \sqrt{a\nu^2\over 2\pi}\exp\left({-a\nu^2\over 2}\right) 
  {{\rm d\,ln}\nu^2\over {\rm d\,ln}m},
\label{nmgif}
\end{equation}
where $\bar\rho$ is the average background density, 
$A=0.322$, $a = 0.707$, $p=0.3$ and $\nu = \delta_{\rm c}/\sigma(m)$ 
with $\delta_{\rm c}\approx 1.686$ and $\sigma^2(m)$ represents the 
variance in the initial density field, when smoothed on the scale 
$R = (3m/4\pi\bar\rho)^{1/3}$, extrapolated to the present time using 
linear theory.  The Press-Schechter formula has $A=0.5$, $a=1$ and 
$p=0$.  Sheth, Mo \& Tormen (2000) argue that whereas the 
Press-Schechter formula is associated with models in which haloes form 
from a spherical collapse, this more accurate formula is associated with 
ellipsoidal collapse.  

What we really need is $n(m|\delta)$.  Unfortunately, the simplest 
approximation to this distribution is not accurate 
(Lemson \& Kauffmann 1999; Sheth \& Lemson 1999): 
$n(m|\delta)\ne (1+\delta)\,n(m)$.  However, a good approximation 
to the actual distribution can be computed following the work of 
Mo \& White (1996):  
\begin{equation}
n(m|\delta)\approx \Bigl[1+b(m)\delta\Bigr]\,n(m),
\label{nmdelta}
\end{equation}
where $b(m)$ depends on the halo mass (Mo \& White 1996) and on the 
shape of $n(m)$ (Sheth \& Tormen 1999).  This approximation is 
accurate provided that $\delta$ is defined by smoothing the density 
field on scales that are sufficiently large that a randomly placed cell 
contains many haloes.  Equations~(10) and~(11) in Sheth \& Tormen (1999) 
give the $b(m)$ relation associated with the $n(m)$ distribution we use 
in this paper (our equation~\ref{nmgif}).  For massive haloes, $b(m)$ 
increases with $m$, so the ratio of the number of massive to less massive 
haloes is larger in dense regions than in underdense regions.  The exact 
form of the predicted dependence on $\delta$ is slightly more complicated 
than the simple approximation shown above, and is in reasonably good 
agreement with simulations (Sheth \& Tormen 2000).  
When computing the model predictions which follow, we actually use 
the exact formula for $n(m|\delta)$, rather than the simpler 
approximation shown above.  

We now turn to the other term, $p(v|m,\delta)$.  
To model $p(v|m,\delta)$ we will assume that the velocity of any 
given dark matter particle is the sum of two terms, 
\begin{equation}
 v = v_{\rm vir} + v_{\rm halo}:
\label{vsum}
\end{equation}
the first is due to the velocity of the particle about the centre of 
mass of its parent halo, and the second is due to the motion of the 
centre of mass of the parent.  We will assume that each of these 
terms has a dispersion which depends on both halo mass and on 
the local environment, so that 
\begin{equation}
 \sigma^2(m,\delta) = 
 \sigma^2_{\rm vir}(m,\delta) + \sigma^2_{\rm halo}(m,\delta).  
\label{sigtot}
\end{equation}
The assumption that the virial motions within a halo are independent 
of the halo's environment, is probably reasonably accurate.  It is not 
clear that the same is true for the halo speeds.  Indeed, in the next 
section we will show that haloes in dense regions move faster than 
those in underdense regions.  It will turn out, however, that the 
fraction of regions in which $\sigma^2_{\rm halo}(m,\delta)$ is 
significantly different from $\sigma^2_{\rm halo}(m,0)$ is quite 
small.  This means that neglecting the density dependence of the second 
term should be a reasonable approximation.  

In what follows, we will assume that the dispersion depends on the 
mass of the parent halo, but not the local density: 
 $\sigma^2(m,\delta) = \sigma^2(m)$.  
In fact, we will make the even stronger assumption that what is true 
of the second moment is also true of the distribution itself:  
$p(v|m,\delta)=p(v|m)$.  In this case, we can rearrange the order of 
the integrals in the numerator of equation~(\ref{fvexact}), and then 
integrate over $\delta$ to get 
\begin{equation}
 f(v) = {\int {\rm d}m\,m\,{n(m)\over\bar\rho}\,p(v|m)},
\label{fv}
\end{equation}
where we have used the fact that the integral over $\delta$ gives 
the universal mass function, and the integral of $m$ times the 
universal mass function over all $m$ gives the average density.  
To proceed, we need a model for the actual shape of $p(v|m)$.  
Since $v$ is the sum of two random variates, we study each in turn.  

Consider the first term, $v_{\rm vir}$.  
We will assume that virialized haloes are isothermal spheres, so 
that the distribution of velocities within them is Maxwellian.  
This is in reasonable agreement with simulations.  
Accounting for the fact that haloes really have more complicated 
density and velocity profiles is a detail which complicates the 
analysis, but not the logic of our argument.  The main reason for 
this is that in the isothermal sphere model, the velocity dispersion 
of particles within the halo is independent of where they are 
within the halo.  

What is the dispersion of this Maxwellian?  
At any given time, the average density of a virialized halo is 
approximately independent of the halo mass.  This means that the 
mass $m$ and size $r$ of a halo are related:  $m/(4\pi r^3/3)$ 
is a constant, typically equal to about 200 times the critical density 
at that time.  This, with the virial requirement that 
$m/r \propto \sigma^2_{\rm vir}$, implies that the dispersion of the 
Maxwellian depends on halo mass:  
$\sigma_{\rm vir}^2(m,a) \propto m^{2/3}$.
The constant of proportionality depends on time $a$ and cosmology, 
and on the exact shape of the density profile of the halo.  We will use 
the relations provided by Bryan \& Norman (1998) to set this constant:  
\begin{equation}
 \sigma_{\rm vir}(m,a) = 476\,g_\sigma\, (\Delta_{\rm nl}\,E^2)^{1/6}
 \left({m\over 10^{15} M_\odot/h}\right)^{1/3}\ {{\rm km}\over {\rm s}},
 \label{sigmavir}
\end{equation}
where $g_\sigma = 0.9$, 
$\Delta_{\rm nl} = 18\pi^2 + 60 x - 32 x^2$ with 
$x = \Omega(a) - 1$, $\Omega(a) = (\Omega_0/a^3)/E^2(a)$, 
$E^2(a) = \Omega_0/a^3 + \Omega_R/a^2 + \Omega_\Lambda$, 
$a=1/(1+z)$, where $z$ is the redshift, so $a=1$ at the present, 
$\Omega_0(a)$ is the ratio of the energy density in matter to the 
critical density, and $\Omega_\Lambda$ denotes the corresponding 
ratio for the energy density associated with a cosmological constant.  
Fig.~\ref{virdelta} below shows that we are safe in assuming that 
this virial term is independent of local environment.  

We turn, therefore, to the second term, $v_{\rm halo}$.  
It will prove more convenient to first study halo speeds after 
averaging over all environments, before considering the speeds as a 
function of local density.  (This parallels the order in which we 
presented the halo mass function and its dependence on density.)  
Consider a halo of size $r$ at the present time.  
Because the initial density fluctuations were small, the particles 
in this halo must have been drained from a larger region $R$ in the 
initial conditions:  $R/r\approx \Delta_{\rm nl}^{1/3}$, 
where $\Delta_{\rm nl}\approx 200$ or so (see text following 
equation~\ref{sigmavir}).  This means, for example, that massive haloes 
were assembled from larger regions than less massive haloes.  
Suppose we compute the rms value of the initial velocities of all the 
particles which make up a given halo.  If we do this for all the 
haloes of mass $m$, then this is similar to computing the rms velocity 
in linear theory, smoothed on the scale $R(m)\propto m^{1/3}$.  

It is well known that the linear theory prediction for the evolution 
of velocities is more accurate than the linear theory prediction for 
the evolution of the density.  Although linear theory provides a useful 
estimate of the initial rms velocities of patches in the initial 
conditions that become haloes at $z=0$, Colberg et al. (2000) showed 
that it is more accurate to assume that the massive haloes are 
associated with peaks in the initial fluctuation field.  
Bardeen et al. (1986) showed that the linear theory rms velocity of 
regions which are peaks when smoothed on the scale $R(m)$ is slightly 
smaller than that of average regions of the same scale, and Colberg et al. 
showed that the peak estimate agrees with the initial rms velocities of 
the haloes in their simulations to within about 20\% or so.  
Colberg et al. only studied the most massive objects in their simulations; 
in what follows, we will show that the linear theory rms velocities of 
peaks provides an accurate estimate of the initial rms velocities of 
haloes of all masses.  

Since we are interested in the rms velocities of haloes at the 
present time (i.e., not in the initial conditions), we must know how 
these velocities evolve.  Colberg et al. (2000) showed that scaling 
the rms velocities of peaks using the linear theory provided an 
underestimate of the actual growth of the rms velocities of the 
massive clusters in their simulations.  They showed that this procedure 
provides velocities which are accurate to within 40\% or so at $z=0$.
While this is good enough for the results presented in the first part of 
this section (we will discuss why shortly), many of the results to follow 
are sensitive to this difference.  Therefore, Section~\ref{evolve} 
studies the reason for this underestimate.  

To summarize: in what follows, we will assume that at the present time, 
the velocities of all haloes, not just the massive ones, are reasonably 
well described by extrapolating the velocities of peaks (smoothed on the 
relevant scale: $R\propto m^{1/3}$) using linear theory.  For Gaussian 
initial conditions, this means that any given value of $v_{\rm halo}$ is 
drawn from a Maxwellian with dispersion $\sigma^2_{\rm halo}(m)$ given by:
\begin{equation}
 \sigma_{\rm halo}(m) = H_0\Omega_0^{0.6}\ \sigma_{-1} \ 
 \sqrt{1 - \sigma_0^4/\sigma_1^2\sigma_{-1}^2} ,
\label{sigmapeak}
\end{equation}
where $H_0$ and $\Omega_0$ are the Hubble constant and the density 
parameter at the present time, 
\begin{displaymath}
 \sigma_j^2(m) = {1\over 2\pi^2} \int {\rm d}k\ 
 k^{2 + 2j}\,P(k)\, W^2[kR(m)] ,
\end{displaymath}
and $W(x)$ is the Fourier transform of the smoothing window.  
For the TopHat in real space that we will use in the remainder of 
this paper, $W(x) = (3/x^3)\,[\sin(x) - x\,\cos(x)]$.  
Notice that the predicted rms velocity depends both on cosmology 
and on the shape of the power spectrum.  The term under the 
square-root arises from the peak constraint---it tends to unity as 
$m$ decreases:  the peak constraint becomes irrelevant for the less 
massive (small $R$) objects.  

\begin{table}
 \begin{center}
  \begin{tabular}{lcccccc}
   Model & $\Omega_0$ & $h$ & $\sigma_8$ & $\sigma_{\rm fit}$ (km/s) & $R_{\rm fit}$ (Mpc/$h$)& $\eta$ \\
    & & & & & & \\
   SCDM  & 1.0 & 0.5 & 0.60 & 513.9 & 15.85 & 0.87 \\
   OCDM  & 0.3 & 0.7 & 0.85 & 474.6 & 34.67 & 0.85 \\
   $\Lambda$CDM & 0.3 & 0.7 & 0.90 & 414.7 & 34.67 & 0.85 \\
  \end{tabular}
 \end{center}
\caption{Model parameters}
\label{params}
\end{table}

We have found that 
\begin{equation}
 \sigma_{\rm halo}(m) = 
 {\sigma_{\rm fit}\over 1 + (R/R_{\rm fit})^\eta}
\label{sigvfit}
\end{equation}
provides a good fit to equation~(\ref{sigmapeak}) in the CDM family 
of models; Table~1 gives the values of $\sigma_{\rm fit}$, $R_{\rm fit}$ 
and $\eta$ for the three representative models we consider in this paper.  
With these values, the fit is better than about 2\% for SCDM, and 
about 1\% for $\Lambda$CDM, over the range $0.05\le R\le $50 Mpc/$h$.

\begin{figure}
\centering
\mbox{\psfig{figure=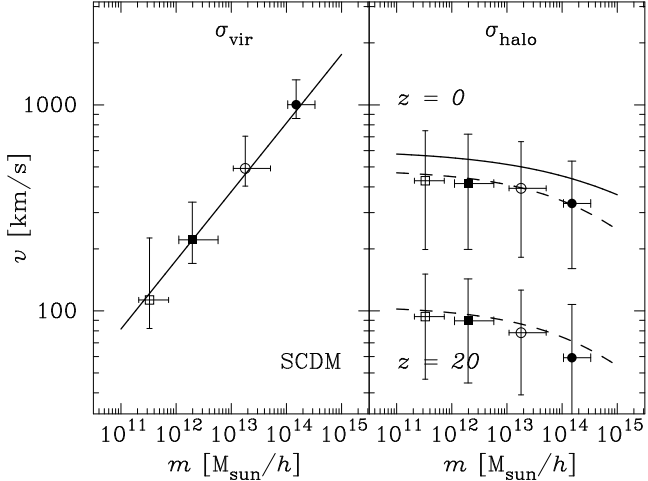,height=6.5cm,bbllx=72pt,bblly=58pt,bburx=629pt,bbury=459pt}}
\caption{Dependence on halo mass of the nonlinear ($\sigma_{\rm vir}$) 
and linear theory ($\sigma_{\rm halo}$) terms in our model.  
Solid curves show the scaling we assume, and symbols show the 
corresponding quantities measured in the $z=0$ output time of the SCDM 
GIF simulation.  Error bars show the 90 percentile ranges in mass and 
velocity.  Dashed curve in panel on right shows the expected scaling 
after accounting for the finite size of the simulation box.  
Symbols and curves in the bottom of the panel on the right show
the predicted and actual velocities at $z=20$.  
}
\label{assumes}
\end{figure}
\begin{figure}
\centering
\mbox{\psfig{figure=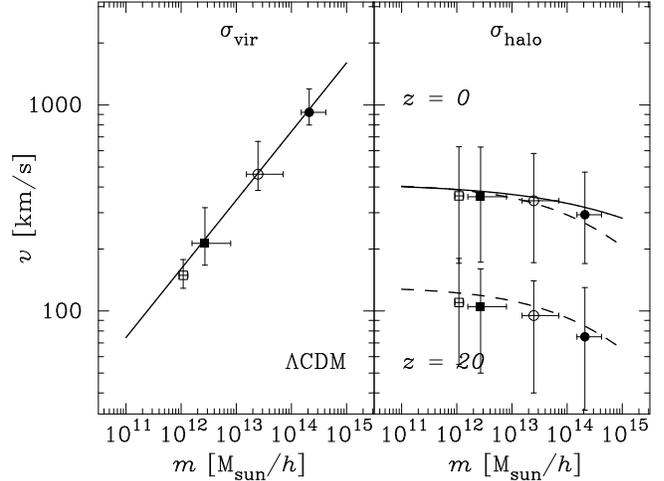,height=6.5cm,bbllx=72pt,bblly=58pt,bburx=629pt,bbury=459pt}}
\caption{As for the previous Figure, but for $\Lambda$CDM.  
Since the simulation box is larger, finite size effects are less dramatic.}
\label{assumel}
\end{figure}

Fig.~\ref{assumes} compares the dependence on mass we assume for 
our two terms with that measured in numerical simulations of clustering 
in an SCDM universe (see Kauffmann et al. 1999 for details of the GIF 
collaboration simulations).  
The symbols with error bars show the median and ninety percentile 
ranges in mass and velocity.  
Open squares, filled squares, open circles and filled 
circles show haloes which have 60-100, 100-$10^3$, $10^3$-$10^4$ and 
$10^4$-$10^5$ particles, respectively.  There are two sets of symbols 
in the panel on the right.  For the time being, we are only interested 
in the symbols in the upper half which show halo velocities at $z=0$.  
The solid curves in the two panels show the scalings we assume.  

Although the scaling of the virial term with mass is quite accurate, 
it appears that the extrapolated linear theory velocities are 
considerably in excess of the measurements in the simulations.  
This is almost entirely due to the finite size of the simulation 
box.  The upper dashed curve shows the effect of using 
equation~(\ref{sigmapeak}) to estimate the rms speeds of halos, 
after setting $P(k)=0$ for $k<2\pi/L$, where $L$ is the box-size:  
$L=85$ Mpc/$h$ for SCDM and it is 141 Mpc/$h$ for $\Lambda$CDM.   
Thus, the two panels show that our simple estimates of the two 
contributions to the variance of $p(v|m)$ are reasonably accurate.  

Fig.~\ref{assumel} shows a similar comparison between our 
model scalings and the GIF simulation of a $\Lambda$CDM cosmology.  
In this case, the simulation box is larger, so the finite box-size 
effects on the halo velocities are much less dramatic.  
Although the accuracy of the scaling of the virial term with mass is 
well known, the reasonably good agreement of the extrapolated linear 
theory peak velocities, over the entire mass range present in the 
simulations, is new.  

Notice that the two terms scale differently with halo mass; indeed, 
to a first approximation, one might even argue that halo speeds 
are independent of halo mass.  
Since our model for the average halo speeds is really only accurate 
to within about 40\% or so (e.g. Colberg et al. 2000 for the massive 
haloes), it is interesting to consider how much we expect our use of 
linear theory peak velocities, and our neglect of the possibility 
that the speeds may depend on environment, to influence the results 
which follow.  
In the $\Lambda$CDM model at small masses, the linear theory prediction
is not affected by box size (the solid and dashed curves in the panel 
on the right are similar), and it provides a good description of 
the average halo speeds.  It is at larger masses where the inaccuracy 
of linear theory may be problematic.  Fortunately, Fig.~\ref{assumel} 
shows that $\sigma_{\rm halo}(m) < \sigma_{\rm vir}(m)$ for massive 
haloes.  Since massive haloes have larger dispersions than less massive 
haloes, the large velocity tail of $f(v)$ is determined primarily by the 
nonlinear virial motions within massive haloes, rather than by the 
peculiar motions of the halo centres of mass.  For this reason, 
the large velocity tails of $f(v)$, at least, are unlikely to be 
sensitive to inaccuracies in our treatment of halo velocities, or to 
our neglect of the possibility that halo speeds may depend on their 
environment.  

We are finally in a position to specify our model for $p(v|m)$.  
Recall that the virial motions are Maxwellian, and that, for 
Gaussian initial density fluctuations, our linear peaks theory 
model of the halo motions means that they too are Maxwellian.  
Thus, in our model, each of the three cartesian components 
of the velocity of a dark matter particle in a clump of mass $m$ is 
given by the sum of two Gaussian distributed random variates, one with 
dispersion $\sigma^2_{\rm vir}(m)/3$ and the other with 
$\sigma^2_{\rm halo}(m)/3$.  If we further assume that the motion 
around the clump centre is independent of the motion of the clump as 
a whole, so these two Gaussian variates are independent, then $p(v|m)$ 
is Maxwellian (equation~\ref{mx} gives the exact shape) with a 
dispersion which is the sum of the individual dispersions 
(the sum in quadrature of equations~\ref{sigmavir} and~\ref{sigmapeak}).  
This model for $p(v|m)$ can be thought of as a simple way in which 
the contributions to the velocity distribution statistic are split up 
into a part that is due to nonlinear effects (the first term) 
and a part which follows from extrapolating linear theory to a  
later time (the second term).  

Our model should be reasonably accurate even if the initial conditions 
were non-Gaussian.  Of course, non-Gaussian initial conditions have 
different $n(m)$ and $n(m|\delta)$ distributions than the ones we use 
here, though they can be computed from the statistics of the density 
fluctuation field similarly to how they were computed in the Gaussian 
case (e.g. Lucchin \& Matarrese 1988; Robinson \& Baker 1999; Sheth 2000).  
Although the virial motions are likely to still be Maxwellian, the fact 
that the linear peak theory velocities studied above were Maxwellians 
is specific to Gaussian initial conditions.  In general, $p(v|m,\delta)$ 
will likely be different from a Maxwellian, although again, in linear 
theory, it too can be computed from the statistics of the initial 
density field.  

\begin{figure}
\centering
\mbox{\psfig{figure=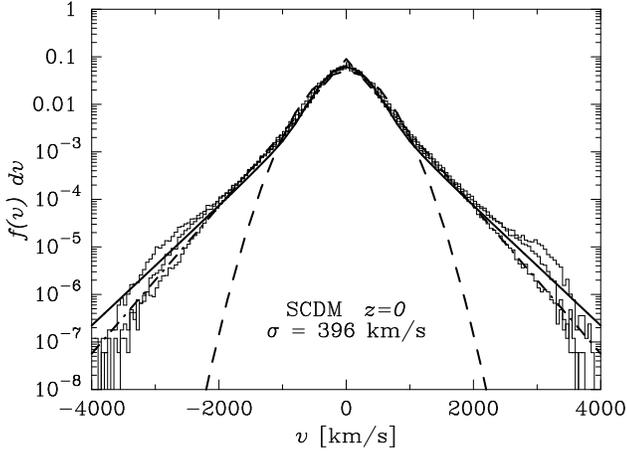,height=6cm,bbllx=72pt,bblly=58pt,bburx=629pt,bbury=459pt}}
\caption{The distribution of one-dimensional peculiar velocities for 
dark matter particles in a SCDM cosmology.  Histograms show the 
distribution of the three cartesian components measured in the GIF 
simulations.  Dashed and dot-dashed curves show Gaussian and exponential 
distributions which have the same dispersion.  The solid curve shows the 
distribution predicted by our model, after accounting for the finite 
size of the simulation box.  The exponential wings are almost entirely 
due to virial motions within haloes.}
\label{pvscdm}
\end{figure}

\subsection{Dark matter particles}\label{fvdm}
\begin{figure}
\centering
\mbox{\psfig{figure=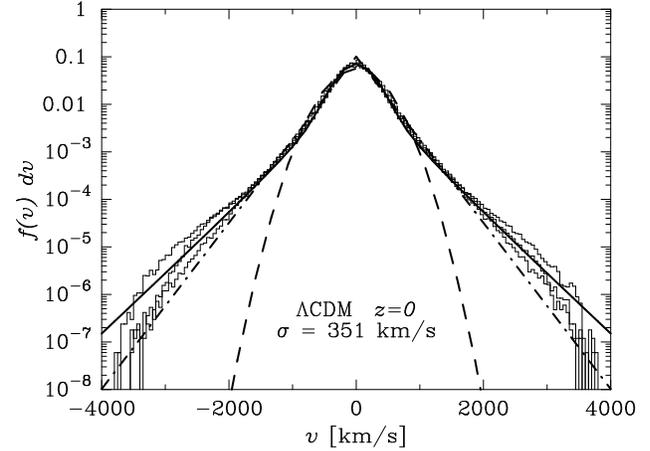,height=6cm,bbllx=72pt,bblly=58pt,bburx=629pt,bbury=459pt}}
\caption{As for the previous Figure, but for $\Lambda$CDM.}
\label{pvlcdm}
\end{figure}

Having described our model assumptions, we are ready to compare 
our predictions with simulations.  In practice, we are only likely to 
observe velocities along the line of sight.  This means that we will 
eventually be interested in the distribution of $f(v)$ projected along 
the line of sight.  Projection changes the Maxwellian $p(v|m)$ 
distributions into Gaussians:
\begin{equation}
p(v|m) = {{\rm e}^{-[v/\sigma(m)]^2/2}\over\sqrt{2\pi\sigma^2(m)}},
\label{pvm}
\end{equation}
where $\sigma^2(m)$ is one third of the sum in quadrature of 
equations~(\ref{sigmavir}) and~(\ref{sigmapeak}).  
Figs.~\ref{pvscdm} and~\ref{pvlcdm} show the one-dimensional $f(v)$ 
distribution for two representatives of the CDM family of models 
presented in Figs.~\ref{assumes} and~\ref{assumel}.
The histograms show the distribution measured in the GIF simulations.
For comparison, the dashed and dot-dashed curves in each panel show 
Gaussians and exponential distributions which have the same dispersion. 
The solid curves show the distribution predicted by our simple model 
(equations~\ref{fv}, \ref{nmgif}, and~\ref{pvm}).  
Exponential wings, and a small $|v|$ core that is more Gaussian than 
exponential are a generic prediction of our model.   
The exponential wings are almost entirely due to nonlinear motions within 
massive haloes, so they are fairly insensitive to our assumptions about 
how fast these haloes move.   
We conclude that our model provides a reasonably good description 
of what actually happens in the simulations.  
It is worth emphasizing that $\sigma(m)$ in equation~(\ref{pvm}) is 
set by the cosmological model and the initial conditions. Thus, the 
second moments of the distributions in Figs.\ref{pvscdm} 
and~\ref{pvlcdm} are not free parameters of our model.  The agreement 
with simulations suggests that our simple treatment of nonlinear and 
linear contributions to the statistic, and our assumption that any 
dependence on local density can be neglected, are sufficiently 
accurate.  

\subsection{Dependence on local density}\label{fvden}
Kepner, Summers \& Strauss (1997) argue that it may be useful to 
compute velocity statistics as functions of local density, where the 
local density is defined as the mass contained within some larger 
smoothing scale, divided by the volume of the larger scale.  
Kepner et al. were actually interested in the pairwise velocity 
distribution, rather than the single particle distribution function 
considered here, but, following Sheth (1996) and 
Diaferio \& Geller (1996), our discussion below of $f(v|\delta)$ applies 
equally to the pairwise statistic in which they were interested.  
Our calculation of $f(v)$ is easily extended to provide a model 
of $f(v|\delta)$.

If we require that 
\begin{equation}
 f(v) = \int f(v|\delta)\,p(\delta)\,{\rm d}\delta
\label{fvdpd}
\end{equation}
where $p(\delta)$ is the fraction of cells which have overdensity 
$\delta$, and we have not written explicitly that both terms in the 
integrand depend on the smoothing scale used to define the local 
density, then equation~(\ref{fvexact}) shows that the density-dependent 
statistic of interest in this section is 
\begin{equation}
 f(v|\delta) = \int p(v|m,\delta)\,{m\,n(m|\delta)\over\bar\rho}\,{\rm d}m.
\label{fvdelta}
\end{equation}
In the context of our model, $f(v|\delta)$ may depend on local density 
either because the distribution of halo masses depends on local density 
or because the distribution of halo speeds depends on the local density, 
or both.  

We have already discussed how the number density of haloes of mass 
$m$ depends on $\delta$ (equation~\ref{nmdelta}), 
We now turn to the other term, $p(v|m,\delta)$.  
Recall that the assumption that the virial term is independent of 
$\delta$ is accurate.  If halo speeds are also independent of their 
environment, then $p(v|m,\delta)$ will be the same sum of two Gaussians 
that we used in the previous section:  $p(v|m,\delta)=p(v|m)$.  
In other words, if halo speeds are independent of their surroundings, 
then $f(v|\delta)$ will differ from the global $f(v)$ simply because 
massive haloes, which have higher virial motions, occur predominantly 
in dense regions.  Similarly, because less dense cells have fewer massive 
halos than average, the virial contribution in underdense cells is 
slightly lower than average.  So, the rms velocity in less dense cells 
should be smaller than in denser cells.  

\begin{figure}
\centering
\mbox{\psfig{figure=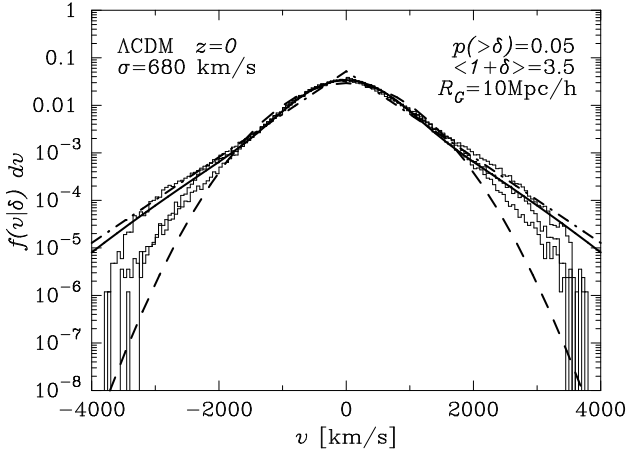,height=6cm,bbllx=72pt,bblly=58pt,bburx=629pt,bbury=459pt}}
\caption{As for the previous Figure, but using only the particles 
which were in the densest regions of a $\Lambda$CDM model.  The 
density was defined by smoothing the density (but not the velocities) 
with a Gaussian filter of size $R_{\rm G}=10$Mpc/$h$.  Figure shows 
results for the densest 5\% of the particles.}
\label{fvdense}
\end{figure}

\begin{figure}
\centering
\mbox{\psfig{figure=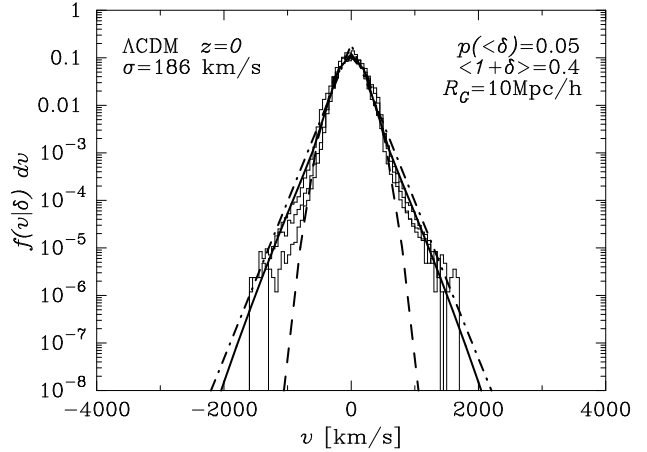,height=6cm,bbllx=72pt,bblly=58pt,bburx=629pt,bbury=459pt}}
\caption{As for the previous Figure, but for underdense regions:  
the least dense 5\% of the particles.}
\label{fvvoid}
\end{figure}

Section~\ref{evolve} shows that, in fact, haloes in denser regions 
move faster than haloes in less dense regions.  Although the shape of 
the distribution of the halo speeds remains approximately Gaussian, the 
dispersion depends on density.  We incorporate this into our model by 
setting 
\begin{equation}
 p(v|m,\delta) = {{\rm e}^{-[3v^2/\sigma^2(m,\delta)]/2}\over
                     \sqrt{2\pi\sigma^2(m,\delta)/3}},
\label{pvmd}
\end{equation}
where 
\begin{equation}
 \sigma^2(m,\delta) = 
 \sigma_{\rm vir}^2(m) + (1+\delta)^{2\mu}\,\sigma_{\rm halo}^2(m),
\label{sigmamd}
\end{equation}
and where the two $\sigma$s are given in equations~(\ref{sigmavir}) 
and~(\ref{sigmapeak}).  As discussed in the next section, 
$\mu$ depends on the cell size on which $\delta$ was defined 
($\mu=0.6$ for the results which follow).  
In the previous subsection, we were able to neglect this dependence 
on $\delta$ because the fraction of regions in which the dispersion 
was significantly different from the average value was small.  Here, 
because we are studying the distribution as a function of $\delta$, 
we cannot ignore this dependence.   

The histograms in Figs.~\ref{fvdense} and~\ref{fvvoid} show the 
$f(v|\delta)$ distribution for the densest and least dense regions 
in the $\Lambda$CDM simulations.  They were constructed by computing
the density field on a $256^3$ grid with the cloud-in-cell technique, 
and smoothing the density field using a Gaussian filter of scale 
$R_G=10$ Mpc/$h$.  Each particle was then assigned a density equal to 
that of the nearest grid point.  The figures show the distribution of 
speeds for the densest and least dense 5\% of the particles.  The 
average overdensity of these particles is given in the top right hand 
corner of the two panels.  

To compute the associated theory curves we must integrate over the 
probability $p(\delta)$ that a cell has overdensity $\delta$.  Since we 
simply wish to illustrate that our model can account for the striking 
differences between the two plots, we will show the result of simply 
using the value of the integrand at the mean value of the overdensity, 
rather than doing the integral exactly.  To compute this representative 
overdensity, we must account for the fact that the theory requires the 
value of the density in a tophat filter whereas the simulations were 
smoothed with a Gaussian filter.  This has two consequences.  First, if 
we require the two filters to contain the same mass, then the scale 
associated with the tophat is about $R_{TH}\approx 1.6\,R_G$.
In addition, we must account for the fact that the mean value of the 
overdensity in such a tophat is different from that in the Gaussian.  
If the distribution of the smoothed $\delta_{\rm G}$ and 
$\delta_{\rm TH}$ fields were Gaussian, then we could compute the average 
$\langle \delta_{\rm TH},R_{\rm TH}|\delta_G,R_G\rangle$.  
This exercise shows that the mean value of the tophat fluctuation 
is typically larger (smaller) than that of the Gaussian in over (under)
dense regions.  This suggests that to compare with theory, we should 
use $R_{\rm TH}=16$ Mpc/$h$, and the associated value of the 
overdensity.  Using $\langle 1+\delta_{\rm TH}\rangle = 3.5$, and 0.4 
for the two cases provides rms values which are similar to those of 
the simulation.  These values in our model give the solid curves in 
the two panels.  The figures show that our model describes the 
differences between the two regimes quite well.  

Including the fact that halo speeds depend on their environment 
(haloes in dense regions move faster than they do in underdense regions) 
was essential to reproduce the simulation results: whereas the dependence 
on density of the halo mass function (equation~\ref{nmdelta}) controls 
the tails of the distribution, it is the dependence of halo speeds on 
density (equation~\ref{pvmd}) which controls the width of the central 
core.  Neglecting this dependence produces distributions which have 
narrower/broader cores than those measured in the dense/underdense 
regions.   Section~\ref{evolve} discusses why halo speeds depend on 
local density.  

\subsection{Dark matter haloes}\label{fvhalo}
Our model can also be used to estimate what happens to the shape of 
$f(v)$ if something other than dark matter particles are used as trace 
particles.  For example, suppose we wish to construct the distribution 
of velocities if only haloes within some mass range are used.  This was 
done by Croft \& Efstathiou (1994).  They found essentially no dependence 
of $\sigma_{\rm halo}$ on cluster mass for the massive clusters in their 
simulations.  They also reported that, in the tails, the distribution 
of massive cluster velocities differed from a Maxwellian.  Our 
model provides a simple explanation for their findings.  

In our model, the distribution of halo speeds is given by 
equation~(\ref{fvexact}), but now each halo contributes only once, so 
the weighting by mass should be removed (both from the numerator and 
the denominator).  Also there should be no contribution from the 
$v_{\rm vir}$ term:  so $p(v|m,\delta)$ is a Maxwellian (a Gaussian 
in one-dimension) with dispersion equal to the second term, 
$\sigma^2_{\rm halo}(m,\delta)$, in equation~(\ref{sigmamd}).  
If the range in halo masses and environments considered is sufficiently 
small, then $f(v)$ will be approximately Gaussian; departures from the 
Gaussian shape will become increasingly apparent as the range in $m$ and 
$\delta$ is increased.  Since $\sigma_{\rm halo}$ depends only weakly on 
halo mass (Figs.~\ref{assumes} and~\ref{assumel}; note that this is 
consistent with Croft \& Efstathiou) any departures from Gaussianity 
are almost entirely due to the dependence of halo speeds on $\delta$.  

As we show in the next section, the halo speeds do depend on $\delta$.  
We were able to neglect this dependence when studying the speeds of dark 
matter particles, because the large velocity non-Gaussian tails were 
determined almost entirely by the virial motions of the particles, and 
not by the halo speeds.  Since these virial motions do not contribute 
to the motions of the haloes themselves, they do not mask the effect 
of the density dependence of halo speeds on the distribution of 
$v_{\rm halo}$.  Thus, in our model, both for the dark matter, and for 
the haloes, the distribution of speeds is predicted to have a non-Gaussian 
tail which is a consequence of adding up Gaussian distributions having 
different dispersions.  However, whereas for the dark matter, this tail 
arises more from the mass dependence of the virial motions within dark 
haloes than from any dependence on environment, the tail in the 
distribution of halo speeds arises more from the fact that the halo 
speeds depend on environment than on halo mass.  

\begin{figure}
\centering
\mbox{\psfig{figure=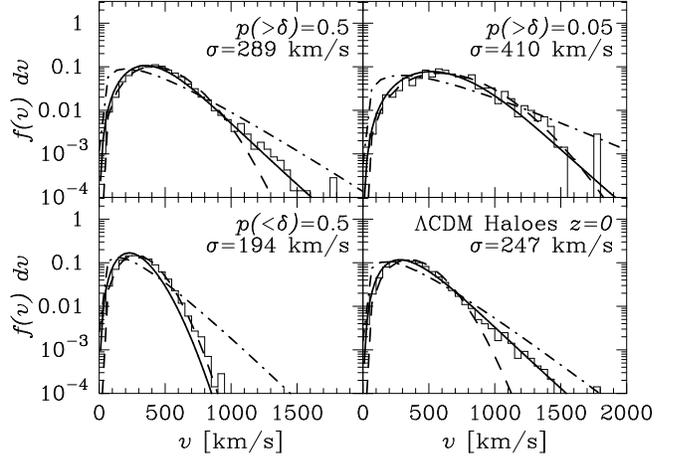,height=6cm,bbllx=72pt,bblly=63pt,bburx=629pt,bbury=452pt}}
\caption{Dependence of the distribution of halo speeds on environment.  
Top left panel shows the distribution of three-dimensional speeds for 
50\% of the haloes, chosen because the local density around them is high, 
and bottom left panel shows the distribution for the other half.  Top 
right panel shows the distribution when only 5\% of the haloes, having 
the highest local density, are used.  Bottom right panel shows the 
distribution of speeds for the total halo sample; it is quite different 
from a Maxwellian (dashed curve).  Dot-dashed curve shows the 
corresponding Expwellian distribution associated with exponential 
one-dimensional distributions.  Solid curves show what our model 
predicts.}
\label{pvhalos}
\end{figure}

To illustrate this, panels on the left of Fig.~\ref{pvhalos} show 
the distribution of halo speeds we obtained after splitting the total 
halo population into two samples, each containing half of the total 
sample, based on the value of the local density around each halo.  
The top left panel shows the distribution for the haloes in the dense 
regions, and the bottom left panel shows the distribution for the haloes 
around which the local density is smaller.  The local density of a halo 
was computed by smoothing the dark matter field onto a $256^3$ grid using 
a Gaussian filter of scale $R_G=10$ Mpc/$h$, as described in the previous 
subsection.  The local density of a halo was then set equal to the 
density of the nearest grid point.  

For comparison, we also show two other distributions: 
a Maxwellian, and an `Expwellian'.  Whereas a Maxwellian has 
independent Gaussian variates for each of the three cartesian 
components of the velocity, the Expwellian has components which are 
exponentially distributed.  The Maxwellian has the form 
\begin{equation}
f_{\rm MB}(v,\sigma) = -2v\ {\partial\over\partial v}\,
{{\rm e}^{-(v/\sigma)^2/2}\over\sqrt{2\pi\sigma^2}},
\label{mx}
\end{equation}
where $\sigma$ is the rms of each of the components, whereas the 
Expwellian is 
\begin{equation}
f_{\rm Exp}(v,\sigma) = -2v\ {\partial\over\partial v}\,
{{\rm e}^{-\sqrt{2}(v/\sigma)}\over\sqrt{2\sigma^2}}.
\label{ex}
\end{equation}
This way of writing the three-dimensional distributions follows from 
the assumption that the vector is drawn in a random direction and its 
length is independent of its direction; this in turn implies that its 
components have identical one-dimensional distributions 
(Feller 1966).

It would appear that the true distribution (histograms) is 
Maxwellian (dashed curves) in the less dense regions, but more 
`Expwellian' (dot-dashed curves) if the local density is higher.  
This is consistent with our model (solid curves), because the range of 
$\delta$ is much smaller in the underdense than the overdense regions.  
To illustrate that the non-Gaussian tail really does arise from mixing 
Maxwellian distributions with different dispersions, and not from 
velocities which are intrinsically non-Gaussian, the top right panel 
shows the distribution of speeds computed using haloes in the denser 
regions, but now with the cut at 5\% rather than 50\% of the total 
sample.  In this case, even though the average value of the local 
density is considerably larger, the range of overdensities is smaller, 
and so our model predicts that the distribution should be well fit by 
a Maxwellian (with a larger dispersion, of course).  The figure 
shows that this prediction is correct: even in these densest regions, 
where the haloes are moving significantly faster than in the rest of 
the simulation box, the components of the velocity are approximately 
Gaussian distributed.  

The figure shows results for the full three-dimensional distributions; 
we did this because the one-dimensional distributions look quite Gaussian 
even when the range in $\delta$ is large.  At first, this may seem 
unexpected:  one usually thinks of the three-dimensional distribution 
as a sum of independent one-dimensional distributions, so if each 
component is Gaussian, then the full three-dimensional distribution 
should be Maxwellian.  The reason why this does not happen is that the 
large $v$ non-Maxwellian tail comes from Gaussians which have large 
dispersions.  If the dispersion is large, then all three components are 
likelier to be large, so the three-dimensional speed is that much more 
likely to be large.  
Since the final distribution we are considering here is a sum over 
Gaussians of different dispersions, the difference between the 
three-dimensional distribution and a Maxwellian appears to be larger 
than the difference between each of the components and a Gaussian.  
Another way of saying this is that the departures from a Gaussian 
distribution are significant when $f(v)\le 10^{-4}$, whereas the 
departures from Maxwellian are significant when $f(v)$ is an order 
of magnitude or so larger.  In effect, this means that it takes fewer 
haloes to measure this difference reliably if one uses the 
three-dimensional distribution rather than the one-dimensional one.  
An important consequence of this is the following.  Suppose that only 
one component of the velocity was measured and found to be approximately, 
but not exactly, Gaussian.  Then the discussion above cautions strongly 
against concluding that the three-dimensional velocities were drawn 
from a Gaussian random field.  

\subsection{On isotropy and correlated components}
There is an interesting aspect of our model which we have not discussed.  
The Maxwell-Boltzmann distribution associated with linear gravitational 
instability theory and Gaussian initial density fluctuations has the 
special property that the three components of the velocity are independent 
Gaussians, and the resulting velocity field is isotropic.  
If the components are non-Gaussian, then isotropy requires that the 
components be correlated (e.g. Feller 1966).  We will show this 
explicitly below for a few toy examples.   

Consider what happens when the velocity is the sum of two rather than 
three components.  We will study the distribution of the angle $\theta$ 
between the vector and the $x$-axis in two steps; we will derive the 
distribution $Q(\tau)$ of $\tau\equiv \tan\theta = v_y/v_x$ first, and 
we will then use the fact that d$\tau/{\rm d}\theta = (1+\tau^2)$ to 
compute the distribution of $\theta$ itself.  And, for simplicity, 
we will only study the distribution in $\theta$ in the regime where 
both $v_x$ and $v_y$ are positive.  (Of course, the distribution in 
the other quadrants can be derived analogously.)  If $v_x$ and $v_y$ 
are independent, but have the same distribution $p$, then 
\begin{equation}
Q(\tau) = \int p(v_x)\, p(v_y = \tau v_x)\,v_x\,{\rm d}v_x.
\end{equation}
If $p$ is Gaussian or exponential, then
\begin{eqnarray}
Q_{\rm Gauss}(\tau) &=& (2\pi)^{-1}\,(1+\tau^2)^{-1} 
\qquad{\rm and}\nonumber\\
Q_{\rm Exp}(\tau) &=& (1/4)\,(1+\tau)^{-2}.
\end{eqnarray}
so that 
\begin{eqnarray}
q_{\rm G}(\theta) &=& 1/2\pi \qquad{\rm and} \nonumber \\
q_{\rm Exp}(\theta) &=& (1/4)\,(\sin\theta+\cos\theta)^{-2}.
\end{eqnarray}
Whereas the distribution of $\theta$ associated with independent 
Gaussians is uniform, the one associated with Exponentials 
is not; it has a minimum at $\pi/4$.  To get a uniform distribution 
in $\theta$, $v_x$ and $v_y$ cannot be independent if they are 
not Gaussian.  Had we chosen $p$ distributions which were less 
centrally peaked than a Gaussian (the exponential is more centrally 
peaked), we would have found $q$ distributions which have maxima, 
rather than mimima, at $\pi/4$.  E.g. a uniform distribution on the 
range $0\le v\le V$ has 
 $q(\theta) \propto \cos^{-2}\theta$ and $\sin^{-2}\theta$ 
for $0\le \theta < \pi/4$ and $\pi/4 \le \theta < \pi/2$, respectively.  

One consequence of this is that if one wishes to model the nonlinear 
velocity field, and one requires that the nonlinear field be isotropic, 
then one's model must have correlations between the components of the 
velocity.  Building such a model is not easy; of course, perturbation 
theory may be helpful here.  
We were able to circumvent this problem by assuming that, for 
sufficiently small ranges in mass and environment, the velocities 
are, in fact, Gaussian.  This allowed us to use all the nice features 
of the Gaussian to build our model relatively easily, without worrying 
about correlations between the different velocity components.  In our 
model, the total velocity distribution is built up by summing over 
many different Gaussian distributions.  Since each of these is isotropic,
the final field is also.  Because the final distribution is non-Gaussian,
but isotropic, the resulting distribution must have correlations between 
$x$, $y$ and $z$ components of the velocity built in, even though we 
never explicitly worried about how to model them correctly.  

We have measured these correlations in the simulations using Spearman's 
rank correlation $\rho$ for the three pairwise permutations of $v_x$, 
$v_y$, and $v_z$ for the total halo population (those in the bottom 
right panel of Fig.~\ref{pvhalos}).  
In all three cases we found values of about $\rho = 0.075$ with 
significance levels (which should lie between zero and one) of about 
$10^{-20}$.  These low values indicate that the correlation is extremely 
significant.  The same test, with two independent normal variates
for the components (same sample size of the halos $N=14089$), yields 
$\rho = 0.0005$ and significance level $0.95$, confirming that 
there is no correlation in this case. 
We then repeated this test using only those halos in the densest, 
and the least dense regions.  Recall that our model assumes that, 
for a sufficiently small range in environment, the distribution should 
be Gaussian.  In both these cases we find that this rank correlation 
statistic suggests that the velocity components are independent and 
that therefore their distribution is, indeed, Gaussian.

\subsection{Galaxies}\label{fvgal}
Galaxies can also be treated within the context of our model, 
provided we assume that all galaxies form within dark matter haloes 
and that, for the most part, any given galaxy contains a small enough 
fraction of the mass of the parent halo that it can be thought of as 
a trace particle within the halo.  This means that all we now need is 
a relation which tells us how, on average, the number of galaxies 
scales with halo mass.  If $N_{\rm gal}(m)$ is known, then we can insert 
it in place of the factor of $m$ in equation~(\ref{fvexact}).  This is 
clearly a very simple assumption, which we will discuss in more detail 
later.  Note that this assumption can also be used to model how different 
the spatial distribution of galaxies is from that of the dark matter 
(Seljak 2000; Peacock \& Smith 2000; Scoccimarro et al. 2000).  
First, we consider the shape of $N_{\rm gal}(m)$.  

One might have thought that, on average, the number of galaxies in 
a halo would be proportional to the available gas, and that the fraction 
of gas should be proportional to the total mass of the parent halo:  
$N_{\rm gal}(m)\propto m$.  However, because gas must cool to form 
stars, and the internal velocity dispersions of haloes increase with 
halo mass (cf. equation~\ref{sigmavir}), gas cools less efficiently 
in more massive haloes.  So, since the number of galaxies actually 
depends on the fraction of gas that can cool, 
$N_{\rm gal}(m)\propto m^\mu$, with $\mu<1$, is more reasonable.  
In addition, there is a lower mass cutoff to this weighting 
scheme, which is associated with the fact that if the potential 
well of a halo is not sufficiently deep, the supernovae formed from 
the first generation of stars may blow the remaining gas out of 
the halo, thus supressing future star formation.  

The fact that $N_{\rm gal}(m)$ depends on $m$ has an important 
consequence for our model.  If there are many galaxies in a halo, then 
it is reasonable to treat each one as a test particle, so that its velocity 
is similar to that of a dark matter particle.  However, mass conservation 
means that a significant fraction of low mass haloes may have at most one 
galaxy within them.  In such haloes, it makes more sense to identify the 
motion of the galaxy with that of the halo.  Even in haloes which contain 
more than one galaxy, there is often one dominant central galaxy.  It 
seems sensible to identify the motion of this galaxy with that of the 
parent halo, so that for this galaxy also, the virial term does not 
contribute.  To model this, we will assume that both the virial and 
halo terms contribute to $N_{\rm gal}-1$ of the galaxies within a halo, 
but that only the halo term contributes for the final remaining galaxy.  

Although this $N_{\rm gal}(m)$ weighting scheme for incorporating 
galaxies into our model is extremely simple, we think it provides a 
useful framework for constructing more complicated, and presumably 
more realistic, schemes.  For example, it is known that ellipticals 
occur preferentially in clusters, whereas spirals occur in the 
field.  Also, within a cluster, ellipticals are more strongly 
concentrated towards the cluster centre than are the spirals.  
The first fact can be incorporated into our scheme simply by using 
two different weighting schemes:  $N_{\rm Ell}(m)$ and 
$N_{\rm Spiral}(m)$.  If ellipticals populate more massive haloes, 
then, since the contribution of massive haloes to the peculiar velocity 
statistic is dominated by the virial term which increases with halo 
mass, one would expect the velocity dispersion of ellipticals to be 
larger than spirals.  

\begin{figure}
\centering
\mbox{\psfig{figure=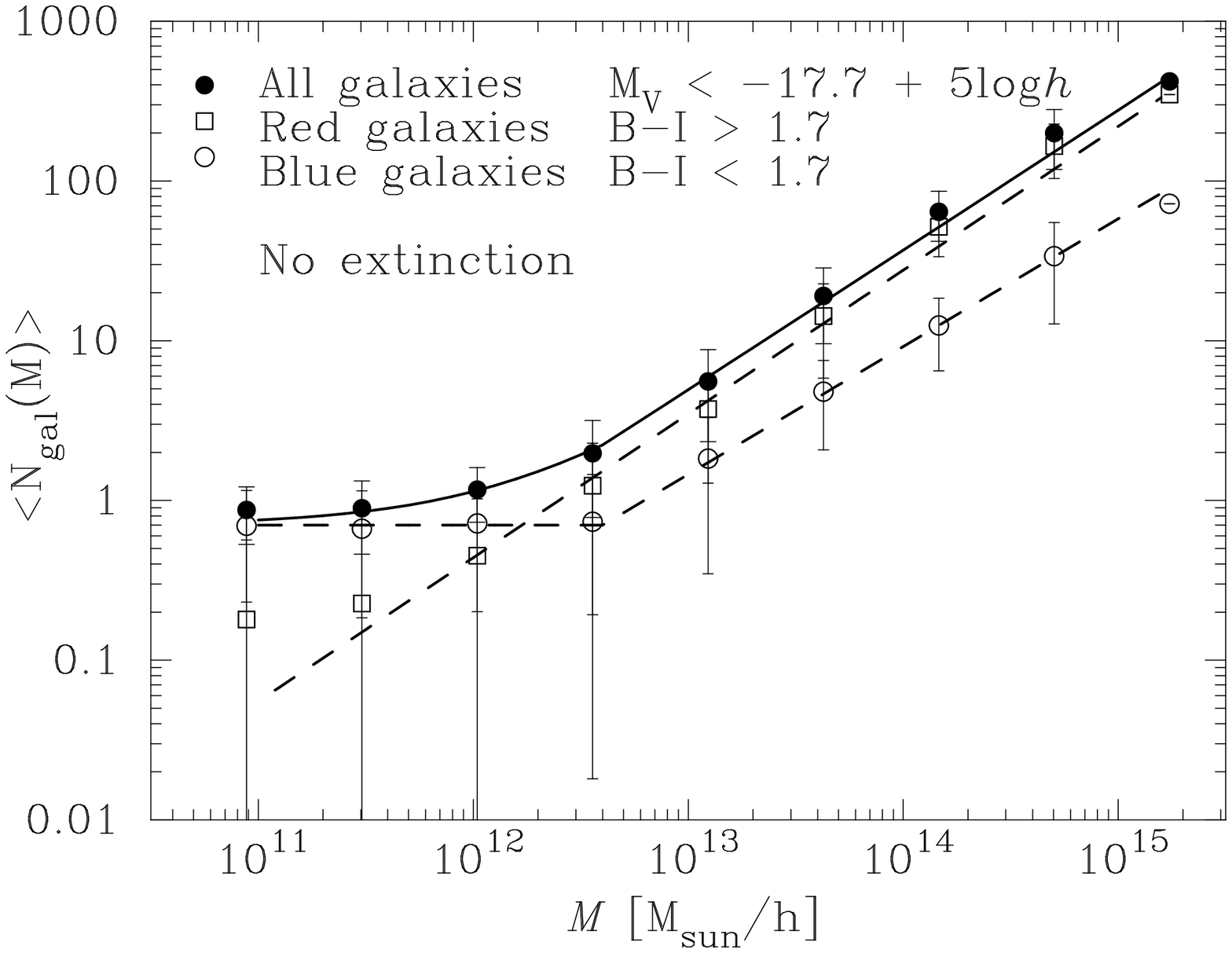,height=6cm,bbllx=72pt,bblly=58pt,bburx=629pt,bbury=459pt}}
\mbox{\psfig{figure=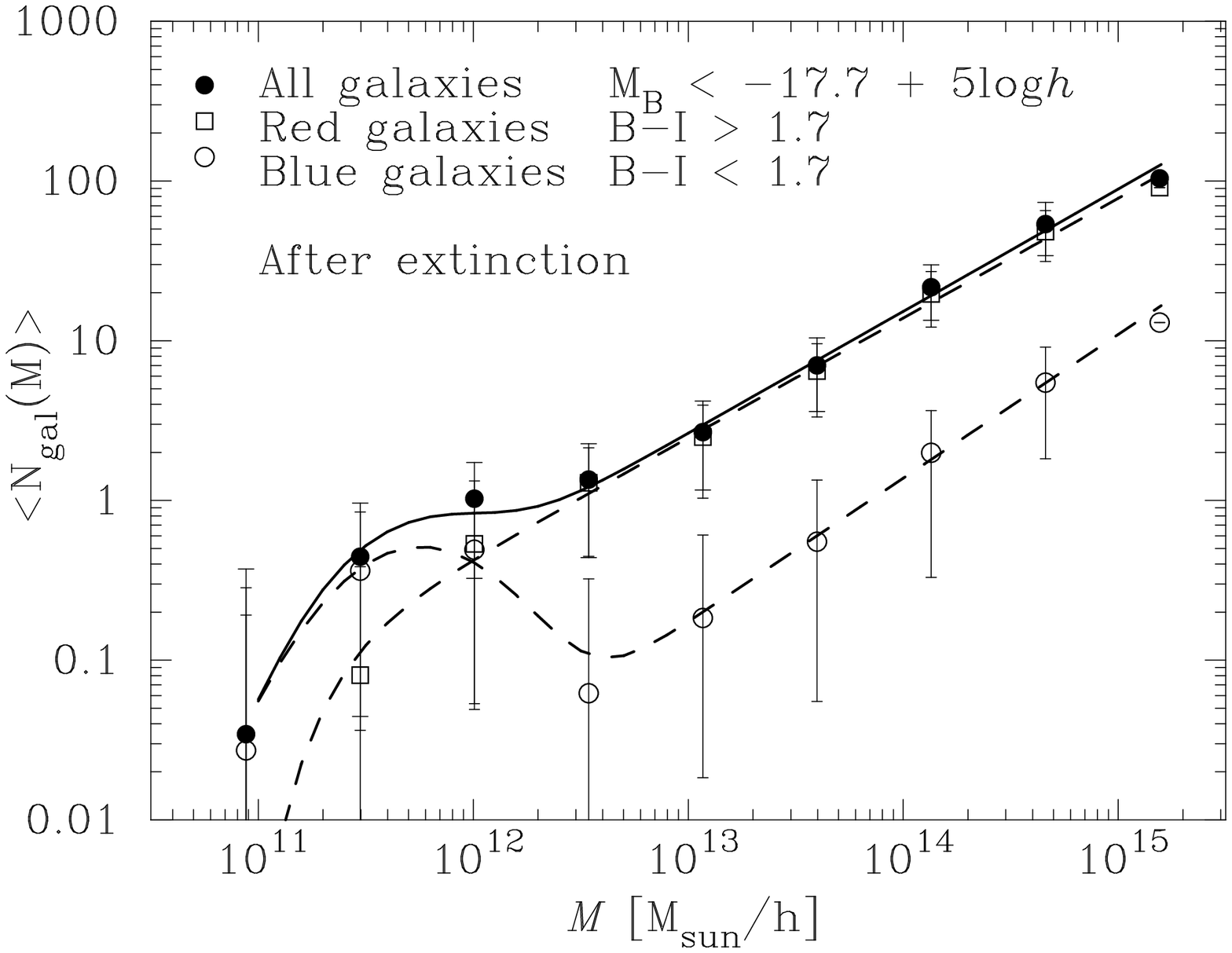,height=6cm,bbllx=72pt,bblly=58pt,bburx=629pt,bbury=459pt}}
\caption{The average number of bright galaxies as a function of parent 
halo mass in the $\Lambda$CDM semi-analytic models of 
Kauffmann et al. (1999).  Top panel shows results for a catalog in which 
no correction for the effects of dust was made.  Bottom panel shows the 
galaxy catalog after correcting for extinction.  Open circles show 
blue galaxies, open squares show red galaxies, and solid circles show 
the sum of the two populations.  Curves show the simple fits to these 
relations discussed in the text. }
\label{ngmh}
\end{figure}

The fact that ellipticals are concentrated more strongly towards cluster 
centres than spirals is slightly more complicated to incorporate.  
If we continue to assume that the galaxies trace the dark matter within 
the halo, then the assumption that all haloes are isothermal spheres 
means that we needn't make any change other than to the weighting term.  
That is to say, because the velocity dispersion is independent of 
position within the isothermal sphere, particles that are nearer the 
centre of the halo move similarly to those that are further away.  
Since, in fact, halo density profiles are more complicated than 
isothermal spheres, the velocity dispersions within the halo are actually 
functions of position, it may be that ellipticals and spirals actually 
have different $p(v|m)$ relations as well.  This would, of course, 
complicate the relation between $\sigma_{\rm Ell}$ and 
$\sigma_{\rm Spiral}$, though it is likely that, in general, 
$\sigma_{\rm Ell} \ge \sigma_{\rm Spiral}$.  


To illustrate how all this works, we will compare our model with 
results from the semi-analytic galaxy formation models of Kauffmann 
et al. (1999).  These models provide the $N_{\rm gal}(m)$ relations 
we need for our model.  Extinction due to dust changes the brightness 
and colours of the model galaxies, so that the $N_{\rm gal}(m)$ relation 
we need depends on whether or not this effect has been account for.  
For this reason we have chosen to study two samples, and they are 
presented in the two panels of Fig.~\ref{ngmh}.
The top panel shows galaxies brighter than $M_V=-17.7 + 5\,\log h$ in a 
model in which the effects of dust have been neglected, and the bottom 
panel shows the $N_{\rm gal}(m)$ relation for galaxies brighter than 
$M_B=-17.7 + 5\,\log h$ after accounting for extinction.  
In each panel, the filled circles show all galaxies brighter than the 
magnitude cut.  The total population can be split into a blue sample 
(open circles) and a red sample (open squares); these are the objects 
which are classified as having colours with $B-I<1.7$ and $B-I\ge 1.7$, 
respectively.  

In what follows, we will use simple approximations (the curves in the 
Figure) to the semi-analytic results.  For the galaxies in the top panel, 
\begin{eqnarray}
N_{\rm Blue}(m) &=& 0.7 \qquad\qquad {\rm if}\ 10^{11}\,M_\odot/h\le m\le M_{\rm Blue}\nonumber \\
                &=& 0.7\,(m/M_{\rm Blue})^{\alpha_B}\qquad {\rm if}\ m>M_{\rm Blue} \nonumber \\
N_{\rm Red}(m) &=& (m/M_{\rm Red})^{\alpha_R} \qquad m\ge 10^{11}\,M_\odot/h 
\nonumber \\
N_{\rm gal}(m) &=& N_{\rm Blue}(m) + N_{\rm Red}(m) ,
\end{eqnarray}
where $M_{\rm Blue} = 4\times 10^{12}\,M_\odot/h$, $\alpha_{\rm B} = 0.8$, 
$M_{\rm Red} = 2.5\times 10^{12}\,M_\odot/h$, and $\alpha_{\rm R} = 0.9$.
Extinction due to dust changes the brightness and colours of the 
model galaxies.  So we have also studied a sample which contains 
galaxies brighter than $M_B=-17.7 + 5\,\log h$, after accounting for 
the effects of dust.  In this case, we find that  
\begin{eqnarray}
N_{\rm Blue}(m) &=& (m/M_{\rm Blue})^{\alpha_B} 
                    + 0.5\,{\rm e}^{-4[\log_{10}(m/10^{11.75})]^2}\nonumber \\
N_{\rm Red}(m) &=& (m/M_{\rm Red})^{\alpha_R}{\rm e}^{-(2\times10^{11}/m)^2} 
\nonumber \\
N_{\rm gal}(m) &=& N_{\rm Blue}(m) + N_{\rm Red}(m) ,
\end{eqnarray}
where $M_{\rm Blue} = 7\times 10^{13}\,M_\odot/h$, $\alpha_{\rm B} = 0.9$, 
$M_{\rm Red} = 3\times 10^{12}\,M_\odot/h$, and $\alpha_{\rm R} = 0.75$.

There are two reasons why one might worry that the semi-analytic models 
underestimate the number of bright galaxies in low-mass haloes.  
The first is that the models only resolve halos with 
$M>1.4 \times 10^{11}\,M_\odot/h$.  Secondly, the semi-analytics include 
a model for dynamical friction which may be more efficient than accurate 
numerical simulations suggest (Springel, White, Tormen \& Kauffmann 2000).  
As a result, galaxies within a halo may merge with the central galaxy 
faster than they should, and so fewer (or just one) bright galaxies 
survive.  On the other hand, in small mass haloes there simply isn't 
enough material available to make a massive satellite object in addition 
to the central galaxy, so having a small number of bright galaxies in 
low-mass haloes is certainly sensible.

\begin{figure}
\centering
\mbox{\psfig{figure=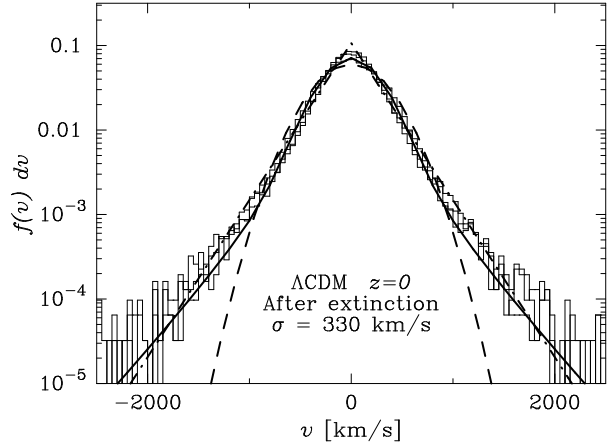,height=6cm,bbllx=72pt,bblly=58pt,bburx=629pt,bbury=459pt}}
\caption{Distribution of peculiar velocities of model galaxies which, 
after correcting for the effects of dust, are brighter than 
$M_B=-17.7+5\log h$.  Histograms show results from the $\Lambda$CDM 
models of Kauffmann et al. (1999).  Dashed and dot-dashed curves show 
Gaussian and exponential distributions which have the same dispersion.  
Solid curve shows the distribution predicted by our $N_{\rm gal}(m)$ 
models (lower panel of previous figure).  }
\label{gifgals}
\end{figure}

\begin{figure*}
\centering
\mbox{\psfig{figure=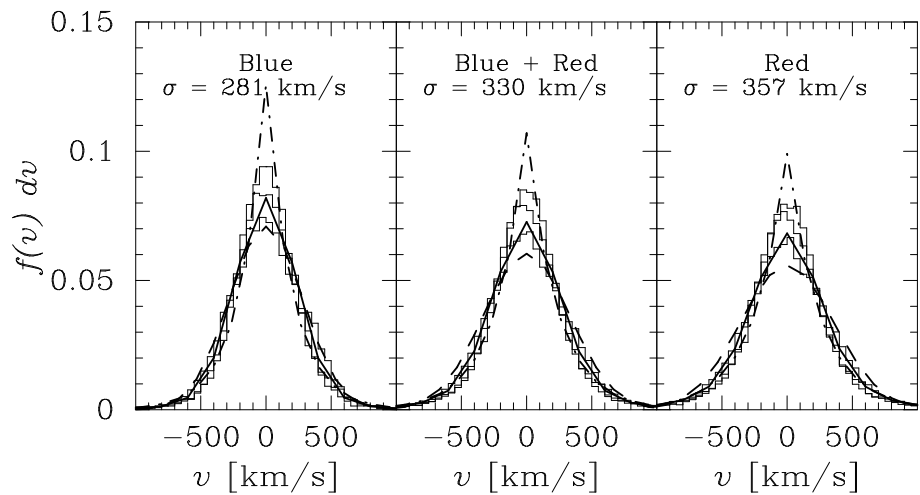,height=6.5cm,bbllx=77pt,bblly=118pt,bburx=618pt,bbury=402pt}}
\caption{Distribution of peculiar velocities of model galaxies as a 
function of colour; colours were assigned after accounting for the 
effects of dust.  The panels on the left and right show blue and 
red galaxies, and the one in the middle shows the sum of the two.  
Line-styles are as in previous figure. }
\label{redblue}
\end{figure*}

Whatever the reason for this low mass cutoff, our main purpose here is 
to show how our model for the dark matter $f(v)$ distribution can be 
extended to describe galaxies.  Thus, although the low mass cutoff in 
the $N_{\rm gal}(m)$ relation may not be in exactly the right place, 
our main interest is in the fact that this cutoff occurs in different 
places for the blue than the red galaxies.  This has quite dramatic 
consequences on the predicted $f(v)$ distribution.  

When Kauffmann et al. (1999) identify galaxies in the simulations, 
each halo with at least one galaxy is assigned a central galaxy. 
The velocity of this central galaxy is chosen to be the same as that 
of the dark matter particle which has the greatest absolute value of 
the potential energy. In the real universe, the central galaxy is 
plausibly almost at rest with respect to the barycentre of its dark 
matter halo, whereas in the simulations the central galaxy moves, 
on average, with a velocity of $\sim 80$ km/s with respect to the 
barycenter of its dark halo.  Therefore, we used the mean velocity of 
the dark particles of the parent halo for the velocity of the central 
galaxy.  The histograms in Fig.~\ref{gifgals} show the resulting 
distribution of bright galaxy peculiar velocities in the 
extinction-corrected simulations (those in the bottom panel of 
Fig.~\ref{ngmh}).  Results for the simulations in which no dust 
correction was applied are sufficiently similar that we decided 
against showing them here.

Notice that although $f_{\rm gal}(v)$ has a slightly different 
shape than the dark matter statistic, it still has tails which are 
significantly different from a Gaussian.  Dashed and dot-dashed curves 
still show Gaussian and exponential distributions which have the same 
dispersion.  The solid curve shows what our simple $N_{\rm gal}(m)$ 
model predicts---it provides a good description of the simulation results.  

Fig.~\ref{redblue} shows how the semi-analytic and our model $f(v)$ 
distributions depend on galaxy type.  Notice that the red galaxies 
have larger dispersions than the blue, as expected if they occur 
predominantly in more massive haloes.  We have chosen to present these 
results using linear plots because peculiar velocity catalogs currently 
available contain on the order of a few thousand galaxies, rather than 
the $10^5$ or so in the simulations.  In practice, the errors on each 
peculiar velocity measurement are likely to be on the order of a few 
hundred km/s.  These will broaden the curves somewhat, making the 
difference between the red and blue samples less dramatic. 

\section{The evolution of halo velocities}\label{evolve}
This section is motivated by the results of Colberg et al. (2000) 
who reported that linear theory underestimates the evolution of 
the speeds of massive haloes.  This underestimate is more severe when 
massive halos have near neighbours. This section argues that, indeed, 
the evolution of halo speeds, and so the accuracy of linear theory, 
depends not so much on halo mass as on local density.  

The previous section showed (Figs.~\ref{assumes} and~\ref{assumel}) 
that combining linear theory with the assumption that haloes at $z=0$ may 
be associated with peaks in the initial conditions provides a reasonably 
good description of the speeds of haloes at $z=0$.  
The lower set of symbols in the panels on the right of 
Figs.~\ref{assumes} and~\ref{assumel} show the halo velocities in the 
simulation at $z=20$.  Following Colberg et al (2000), a halo's velocity 
at $z=20$ is computed using the same particles that are in it at $z=0$, 
but using the velocities they had at $z=20$, rather than the velocities 
they actually have at $z=0$.  The lower dashed curve in the panel on the 
right was computed using equation~(\ref{sigmapeak}) at $z=20$, assuming 
that there was no power in modes with $k \le 2\pi/L$.  
This means that the lower dashed curve differs from the upper one by the 
linear theory velocity growth factor ($\propto \sqrt{a}$ in SCDM); on a 
log plot the two dashed curves have the same shape.  
(Although there is a slight offset---the theory curve should be multiplied 
by about 0.9 if it is to pass through the $z=20$ symbols---it has the same 
shape as the symbols.  Though they do not remark on it, a comparison of 
columns 9 and 17 in Table~2 of Colberg et al. 2000 shows the same offset.)  

Although the dashed curves appear to have the same shape as the symbols, 
both at $z=20$ and at $z=0$, there is a systematic discrepancy: 
the model appears to underestimate the amount by which the speeds of 
massive haloes have evolved.  This discrepancy is the same as that 
reported by Colberg et al. (2000).  
At face value, this suggests that the accuracy of the linear theory 
evolution of halo speeds depends on halo mass.  In fact, we will argue 
that the actual evolution depends not so much on halo mass as on local 
density.  It is well known that the evolution of clustering within dense 
regions is accelerated relative to the average 
(e.g. Tormen \& Bertschinger 1996; Cole 1997)---we show that this applies 
to the evolution of halo speeds also.  In particular, we argue that the 
evolution of halo speeds is best estimated by using an $\Omega_{\rm eff}$ 
which depends on (some suitably defined) local density, rather than the 
global value $\Omega_0$.  We describe a simple model in which massive 
haloes populate denser regions, so that the dependence on density appears 
as a dependence on halo mass.  

After studying the accuracy of the predicted linear theory amplitude, 
this section turns to the direction of motion predicted by linear 
theory.  It shows that nonlinear kicks to a particle's motion may 
substantially change the particle's direction.  Although the amplitude of 
the nonlinear kicks are larger for particles which end up in massive 
haloes, these kicks are random in direction, so they can be estimated 
easily.  The directions of halo motions, on the other hand, are extremely 
well described by linear theory.  That is to say, haloes identified 
at any given time are moving in essentially the same direction as 
predicted by linear theory.  This provides additional justification 
for our model in which nonlinear effects are modelled as arising 
primarily from virial motions within haloes.

\subsection{Massive haloes populate denser regions}
Suppose one assumes that massive haloes form near peaks in the initial 
fluctuation field, and that they are close to local minima in the initial 
potential field (Kaiser 1984; Bardeen et al. 1986).  Colberg et al. (2000) 
show this is true for the most massive objects present at $z=0$.  
In such a model, less massive haloes fall towards the massive ones, 
so one might expect the speeds of less massive haloes to be higher in 
regions which contain massive haloes.  This leads to the question: 
Do massive haloes populate denser regions than average?    

Equation~(\ref{nmdelta}) for $n(m|\delta)$ suggests that massive haloes 
occur predominantly in denser cells (recall that for massive haloes 
$b(m)$ increases with $m$).  Although this is intuitively obvious---one 
reason a cell is denser is because it contains massive objects---the Appendix 
provides a simple quantitative model of the average density of cells 
which are known to contain haloes of a given mass $m$.  Therefore, in a
 model in which less massive haloes stream towards the more massive ones, 
one might expect the speeds of less massive haloes to be higher in denser 
regions than in less dense ones.  Moreover, since the evolution of 
clustering is accelerated in dense regions relative to the average, one 
might expect halo speeds to also evolve faster in dense regions than in 
underdense regions.  

In such a model, massive haloes populate dense regions, so the evolution 
of their speeds is best estimated by using an $\Omega_{\rm eff}$ 
which is larger than the global value.  Moreover, massive haloes which 
have massive near neighbours are, on average, in even denser regions.  
Since $\Omega_{\rm eff}$ is even larger for the massive haloes which 
have massive neighbours, one might expect the discrepancy with the linear 
theory evolution to be even larger for such haloes than for those which 
are relatively isolated.  This is qualitatively consistent 
with what Colberg et al. (2000) found in their simulations:
the discrepancy between the linear theory predicted velocity at $z=0$ 
and the actual velocity of a halo at that time is larger for massive 
haloes which have massive neighbours nearby.  

This argument applies to less massive haloes also, although in this 
case things are complicated by the fact that less massive haloes occupy 
regions spanning a wider range of densities---their speeds evolve 
faster or slower than average depending on whether or not they are 
in denser or less dense regions.  On average, however, less massive 
haloes occupy regions of about average density, and so using the global 
value of $\Omega_0$ to estimate the evolution of the population as a 
whole will be reasonably accurate.  This is why using the global value 
of $\Omega_0$ to extrapolate from $z=20$ to $z=0$ works well for less 
massive haloes, but results in an underestimate of the present day 
velocities of massive haloes.  

How does such a model compare with what happens in simulations?  

\subsection{Comparison with simulations:  speed}
\begin{figure}
\centering
\mbox{\psfig{figure=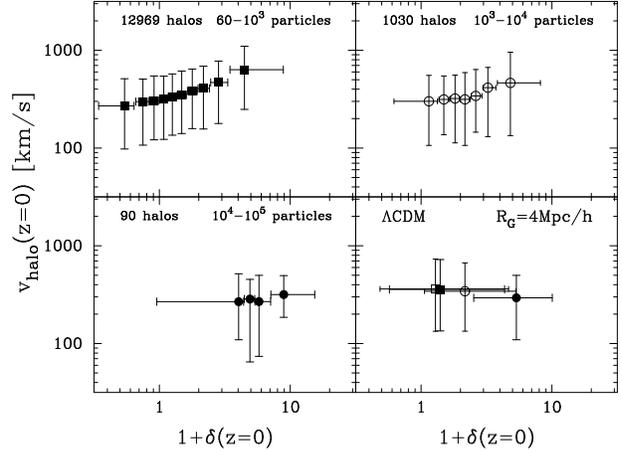,height=6cm,bbllx=78pt,bblly=62pt,bburx=608pt,bbury=452pt}}
\caption{Dependence of halo speed at $z=0$ on local density at $z=0$ 
for haloes of different masses.  Haloes in dense regions are moving 
faster than haloes in underdense regions.  On average, massive haloes 
populate denser regions, and move slightly more slowly than less massive 
haloes.  This is consistent with a model in which massive haloes are 
associated with minima in the initial potential field, and less massive 
haloes are falling towards these minima.}
\label{v0d0r4}
\end{figure}
Fig.~\ref{v0d0r4} shows the velocities of haloes at $z=0$ as a function
of the `local' density at that time.  The local density was computed 
by smoothing the density field at $z=0$ with a Gaussian filter of radius 
$R_G=4$Mpc/$h$.  In the top two and bottom left panels, the bins in the 
x-direction were chosen to contain equal numbers of haloes, the error 
bars show the range containing 90 percent of the haloes in the bin, 
and the symbols show the median velocity and density for each bin.  
The three panels show haloes of different mass ranges.  
In all three panels, there is a clear trend for haloes in dense regions 
to move faster.  Also, comparison of the three panels shows that, on 
average, massive haloes populate denser cells.  The bottom right panel 
shows this trend with halo mass more clearly.  The open squares, filled 
squares, open circles and filled circles show the median velocities of 
haloes, and the median densities of the cells populated by haloes 
containing 60-100, $10^2$-$10^3$, $10^3$-$10^4$ and $10^4$-$10^5$ 
particles, respectively (the same symbols and ranges as in 
Fig.~\ref{assumel}).  The error bars show the range in which 90\% of 
the haloes in the given mass ranges lie (though we have not done so 
here, we could have estimated these ranges directly from the work of 
Sheth \& Lemson 1999).  At $z=0$, massive haloes clearly populate denser 
regions, and they clearly move slightly slower than less massive haloes.  

We have also constructed such a plot using $R_G=10$Mpc/$h$.  
On this larger smoothing scale also, the trend for higher halo speeds 
in denser cells at $z=0$ remains:  
$v\propto (1+\delta)^\alpha$ with $\alpha\ge 0$.  
Although $\alpha$ increases with $R_G$ because the range in 
densities is smaller at large $R_G$, whereas the halo velocities do 
not depend on $R_G$, most of this difference in slope can be accounted 
for as follows.  

\begin{figure}
\centering
\mbox{\psfig{figure=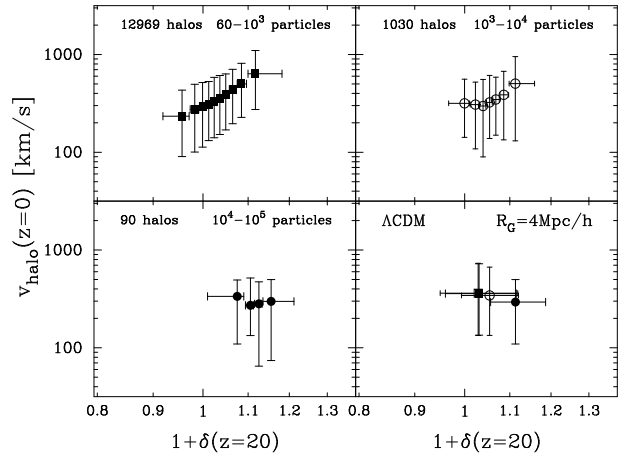,height=6cm,bbllx=78pt,bblly=62pt,bburx=608pt,bbury=452pt}}
\caption{Dependence of halo speed at $z=0$ on local density at $z=20$, 
for haloes of different masses.  At the present time haloes which were 
initially in dense regions are moving faster than haloes which were 
initially in underdense regions.  This trend is true for haloes of all 
masses.  }
\label{v0d20r4}
\end{figure}
\begin{figure}
\centering
\mbox{\psfig{figure=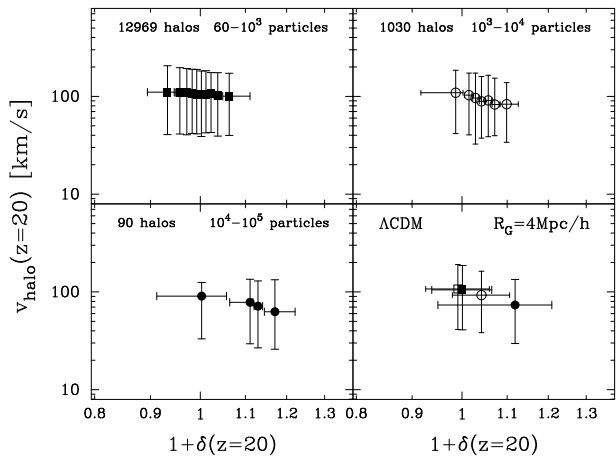,height=6cm,bbllx=78pt,bblly=62pt,bburx=608pt,bbury=452pt}}
\caption{Dependence of halo speed at $z=20$ on local density at $z=20$, 
for haloes of different masses.  Except for a weak trend for haloes in 
dense regions to move more slowly (the trend is reversed at $z=0$ anyway), 
there is little correlation between halo speed and surrounding density.  
This suggests that the tendency to move faster in denser regions, shown in 
the previous figure, must be due to evolution. }
\label{v20d20r4}
\end{figure}

On average, dense cells at $z=0$ were also dense initially.  
The spherical collapse model allows us to estimate the 
initial overdensity $\delta_0$ given the density $1+\delta$ today 
(Mo \& White 1996 provide a simple fitting formula for this relation).  
If we were to rescale the x-axes to show the initial $\delta_0$, 
rather than the density today, then this would move the bins with 
$\delta > 0$ to the left, and the bins with $\delta \le 0$ to the 
right.  The magnitude of the shift would be larger for large 
values of $|\delta|$.  As a result, the plot for $R_G=10$Mpc/$h$
would remain essentially unchanged (this just says that large regions 
have not changed their size very much since the initial conditions, 
and $\delta_0\approx \delta$ for small $\delta$) whereas the plot for 
$R_G=4$Mpc/$h$ would change significantly.  If we were to apply this 
rescaling, then $v\propto [1+\delta_0(\delta)]^\alpha$,
with the nearly the same value of $\alpha$ for a range of 
smoothing scales:  $4\le R_G\le 10$Mpc/$h$.  Of course, once the 
smoothing scale on which $\delta$ is defined becomes very large, the 
trend with `local' density vanishes.  This scale is approximately the 
same as that on which the mass function of haloes approaches the global 
value: $n(m|\delta)\to n(m)[1 + b(m)\delta]$, but $\delta\ll 1$.  

Rather than apply this spherical collapse $\delta_0(\delta)$ rescaling,
we have done the following.  We have taken all the particles that are 
in a halo at $z=0$, found their positions at $z=20$, and plotted 
$v_{\rm halo}(z=0)$ versus $1+\delta$ at $z=20$.  Fig.~\ref{v0d20r4} 
shows this when the smoothing scale used to define the `local' density 
at $z=20$ was $R_G=4$Mpc/$h$.  Notice the correlation between higher 
speeds and local densities.  This figure should be compared with 
Fig.~\ref{v20d20r4} in which the local densities were computed using 
$R_G=4$Mpc/$h$ at $z=20$ but the halo velocities were computed using 
the velocities the particles had at $z=20$.  The difference between 
the two plots is striking.  Notice that, for all but the most massive 
haloes, the velocities at $z=20$ show no dependence on the local density 
at that time.  Because a sphere of size $R_G=4$Mpc/$h$ in which the 
overdensity is $\delta$ at $z=0$ was actually larger at $z=20$ [by a 
factor of $(1+\delta)^{1/3}$], it is not really fair to compare plots 
in which the same smoothing scale is used for both times.  
We have found no dependence of $v_{\rm halo}(z=20)$ on local density 
at $z=20$ even when the density is computed by smoothing with larger 
filters.  (At $R_G=20$Mpc/$h$, there is a weak trend for the halo speeds 
to be smaller in denser cells, even for the less massive haloes.  
It would be interesting to know if this were real.  Though we have 
not done so, it should be possible to do the linear theory calculation 
for the predicted bulk motion of a region subject to the condition that 
the region of interest is embedded in a larger scale over/underdensity.  
For the special case in which the scales on which the initial velocities 
and densities are smoothed are the same, linear theory predicts no 
dependence of the initial velocities on the initial densities, which 
is approximately what our bottom panel shows.)  
We interpret the fact that halo velocities at $z=20$ are independent 
of local density at $z=20$, but halo velocities at $z=0$ do depend on 
local density at $z=20$ as strong evidence that the evolution of halo 
velocities depends on their environment.  

\begin{figure}
\centering
\mbox{\psfig{figure=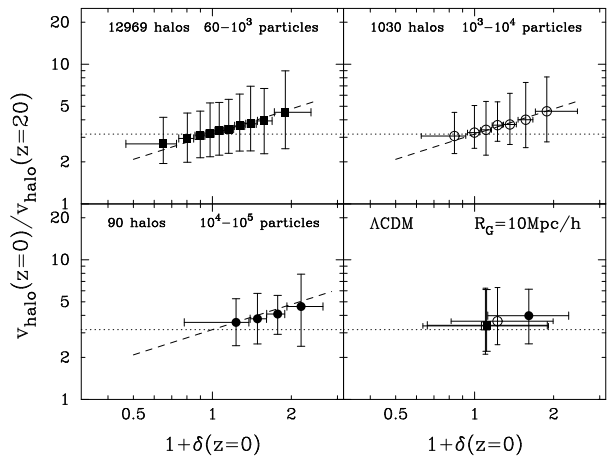,height=6cm,bbllx=87pt,bblly=62pt,bburx=608pt,bbury=452pt}}
\caption{Dependence of the ratio of the halo speeds at $z=0$ and 
$z=20$ on local density at $z=0$, for haloes of different masses.  
Dotted lines in each panel show the linear theory prediction; 
dashed lines, the same in each panel,  show equation~(\ref{vgrowthfit}).  
The measured ratio is higher in denser regions; velocities in dense 
regions evolve faster than in less dense regions.  The dependence on 
density is approximately the same for all haloes---it does not depend 
on halo mass.}
\label{v0v20r10}
\end{figure}

To quantify this, Fig.\ref{v0v20r10} shows the ratio of the median halo 
velocity at $z=0$, $v_0$ to that at $z=20$, $v_{20}$, as a function of 
local density at $z=0$, computed using $R_G=10$Mpc/$h$.  
For the $\Lambda$CDM model studied here, the linear theory prediction 
is that $v_0/v_{20} = 3.16$; 
it is shown as the horizontal dotted curve in the panels.  The different 
symbols in each bin show the different mass ranges.  This shows that 
the evolution of halo velocities depends more on environment 
than on halo mass.  The dashed curves show 
\begin{equation}
 {v_0\over v_{20}} = (1+\delta)^{\mu(R)}\,f_{\rm lt},
\label{vgrowthfit}
\end{equation}
where  $f_{\rm lt}\propto\Omega_0^{0.6}$ is the growth factor predicted by 
linear theory, and $\mu(R=10\,{\rm Mpc}/h)=0.6$.  The dashed curves in 
the three panels all show this same scaling---the dependence of halo 
speed on local density is independent of halo mass.  The trend in the 
bottom right panel of Fig.\ref{v0v20r10} only reflects the fact that 
massive halos are in denser regions.

\begin{figure*}
\centering
\mbox{\psfig{figure=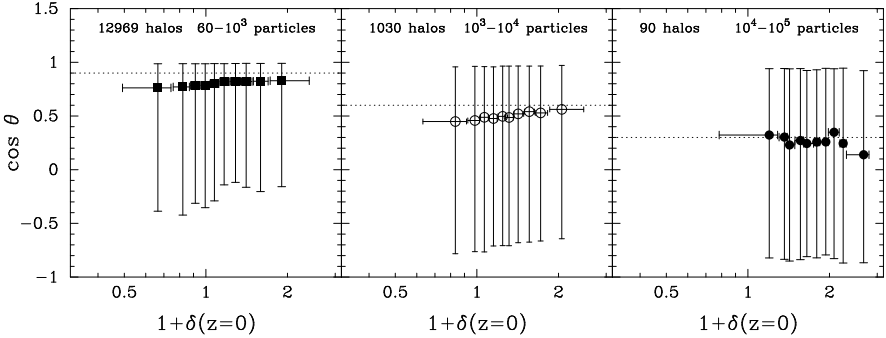,height=4.5cm,bbllx=17pt,bblly=174pt,bburx=759pt,bbury=458pt}}
\caption{The dot product of the linear and actual directions of motion 
for dark matter particles in less massive (left), intermediate 
(middle) and massive (right) halos, as a function of overdensity.
Symbols show the median and error bars show the range containing 
90\% of the particles in each bin.  The nonlinear kicks to a particle's 
motion are larger in massive halos.  Dashed line shows the expected 
median if kicks are randomly distributed. }
\label{v0dotv20}
\end{figure*}

As discussed earlier, $\mu(R)$ depends on the scale $R$ on which the 
local density field was defined.  A reasonable approximation to the 
dependence on $R$ can be got by setting 
 $\mu = 0.6\,\sigma^2(R)/\sigma^2(10\,{\rm Mpc}/h)$, where 
$\sigma^2(R)$ denotes the variance in the present day density field 
when smoothed on scale $R$.  Alternatively, as discussed earlier, 
one might have chosen to incorporate the dependence on $R$ by using 
$[1+\delta_0(\delta)]$ instead.  
The main point we wish to make is that our simple fit, 
equation~(\ref{vgrowthfit}), strongly suggests that the dependence 
of the evolution of halo speeds on local density can be very simply 
parametrized by replacing the global value of $\Omega_0$ with what is, 
effectively, a local one.  

Because the rms speeds of halos depend on environment, one might have 
wondered if the full distribution of halo speeds, rather than just 
the rms, depends on environment.  However, as we showed in the 
previous section (Fig.~\ref{pvhalos}), if one studies a sufficiently 
small range of halo masses and environments, then 
$p(v_{\rm halo}|m,\delta)$ is Gaussian/Maxwellian, to a good 
approximation.  When the range in $m$ and $\delta$ is increased, the 
distribution develops a non-Gaussian large $v$ tail; this happens not 
because the velocities themselves are non-Gaussian, but because, when 
one considers a wide range in $\delta$, one sees the effect of summing 
up Gaussians which have a wide range of widths.

\subsection{Comparison with simulations:  direction}
In addition to a speed, linear theory also predicts a direction of 
motion.  This subsection shows that the linear theory direction 
does not provide a good indicator of the direction of motion of 
dark matter particles at $z=0$.  It is, nevertheless, a very good 
indicator of the direction of halo motions.  Essentially, this is 
because the motion of a dark matter particle receives substantial 
nonlinear kicks as it becomes part of a virialized halo---the 
amplitude of the nonlinear kicks increases with halo mass.  
Haloes themselves do not suffer substantial kicks to their motion.  

In what follows, we will interpret the difference 
 ${\boldmath v}_{0}-{\boldmath v}_{20}$ as the nonlinear kick to a 
particle's motion.  Fig.~\ref{v0dotv20} shows 
\begin{equation}
\cos\,\theta = {{\boldmath v}_{0}\cdot {\boldmath v}_{20}\over
|{\boldmath v_0}||{\boldmath v_{20}}| }
\end{equation}
as a function of local density for the dark matter particles for a 
range of halo masses.  The figure clearly shows that the nonlinear 
kicks to a particle's motion are larger in massive haloes.  
If these kicks were randomly distributed in direction, 
then the present day direction would be the sum of two independent 
Gaussian vectors having different, but known, dispersions,   
$\sigma^2_{\rm halo}(m)$ and $\sigma^2_{\rm vir}(m)$, which can 
estimated from the results of the previous section (Fig.~\ref{assumel}).
This allows us to compute the distribution of $\cos\theta$.  
For random kicks the mean value should be 
\begin{equation}
\langle \cos\,\theta\rangle =  {1\over \sqrt
{1 + \sigma^2_{\rm vir}(m)/\sigma^2_{\rm halo}(m)}}.
\end{equation}
(This follows from the fact that if the kicks are random, then the initial 
direction of motion and the direction of the kick are, on average, at right 
angles to each other.)  For the range of halo masses in the three panels 
above, this would require $\langle\cos\theta\rangle\approx 0.9$, 0.6 and 
0.3, which is in reasonable agreement with the Figure.  

\begin{figure*}
\centering
\mbox{\psfig{figure=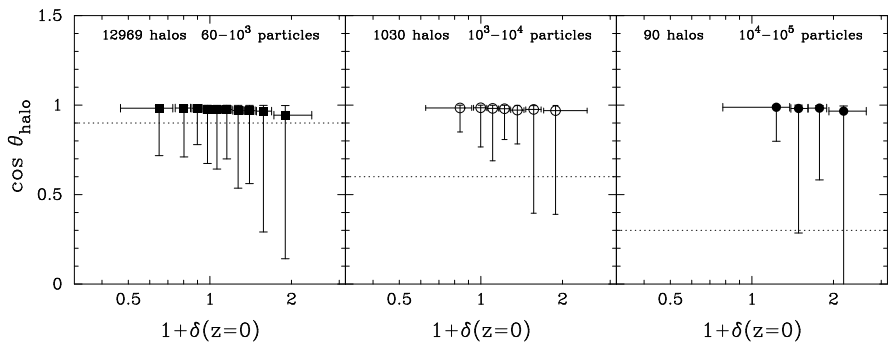,height=4.5cm,bbllx=17pt,bblly=174pt,bburx=759pt,bbury=458pt}}
\caption{The dot product of the linear and actual directions of motion 
for less massive (left), intermediate (middle), and massive 
(right) halos, as a function of overdensity.  Symbols show the 
median and error bars show the range containing 90\% of the haloes
in each bin.  The nonlinear kicks to a halo's motion are small; 
they are larger in denser regions.  
Dotted lines (same as the previous figure) represent the typical kick 
received by a dark matter particle. }
\label{v0dotv20h}
\end{figure*}

\begin{figure*}
\centering
\mbox{\psfig{figure=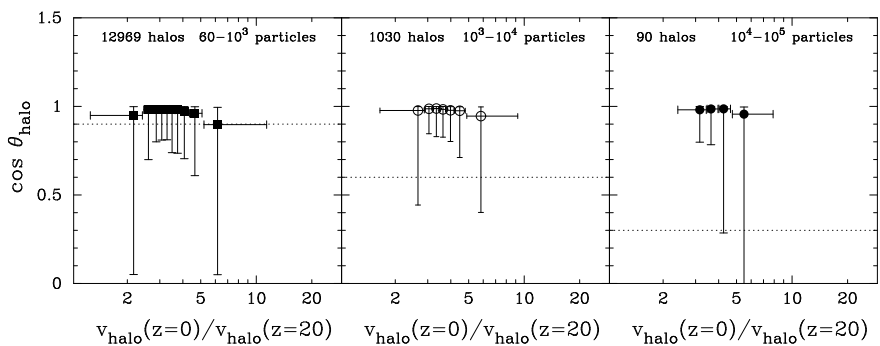,height=4.5cm,bbllx=17pt,bblly=171pt,bburx=759pt,bbury=458pt}}
\caption{Relation between the direction of a halo's motion and the 
evolution of its speed.  Panels show the dot product of the linear and 
actual directions of motion for less massive (left), intermediate (middle) 
and massive haloes (right) versus the ratio of the actual and 
initial, linear, speeds.  The large scatter in directions at small and 
large $v_0/v_{20}$ are from haloes in dense regions.  }
\label{thetavsv}
\end{figure*}

Fig.~\ref{v0dotv20h} shows the corresponding distribution for the halo 
motions.  Although the nonlinear kicks to a halo's motion are larger in
denser regions, these kicks are almost always smaller than the initial 
velocity.  

Fig.~\ref{thetavsv} presents one consequence of the fact that the 
nonlinear kicks to a halo's motion are larger in denser regions.  The 
figure shows that the deflection $\theta_{\rm halo}$ is large both at 
small and large values of $v_0/v_{20}$.  Because large values of 
$v_0/v_{20}$ are associated with dense regions (Fig.~\ref{v0v20r10}), 
the large range in $\theta_{\rm halo}$ at large $v_0/v_{20}$ is, perhaps 
expected.  The reason this happens at small $v_0/v_{20}$ as well is 
slightly more subtle.  There is a hint associated with the fact that 
the magnitude of the scatter is the same in the smallest and largest 
bins, and it is approximately the same as the scatter in the largest 
density bin in Fig.~\ref{v0dotv20h}.  Presumably, the haloes with small 
$\cos\theta_{\rm halo}$ and small $v_0/v_{20}$ are precisely the ones 
which scatter low in the densest cells in Fig.~\ref{v0v20r10}; these 
are the haloes in dense regions which have suffered large kicks to their 
initial velocities, so they are now travelling in quite different 
directions than they were initially.  Of course, because these are the 
haloes in the densest bins, they are only a small fraction of the set 
of all haloes.  Figs.~\ref{v0dotv20h} and~\ref{thetavsv} show that, for 
most haloes, the linear theory direction is a good indicator of the 
direction of motion today.  

To summarize:  This section makes two points.  
The first is that linear theory provides a good description of the 
initial speeds of haloes, but tends to underestimate the rate at which 
the speeds of massive haloes evolve.  Although second-order perturbation 
theory is the best way to study why this happens, our results suggest that 
this happens because the evolution of halo speeds is faster in denser 
regions.  The speeds of massive haloes evolve faster than linear theory 
predicts primarily because they populate denser regions.  Therefore, it 
may be sufficient to model the evolution of halo speeds by replacing the 
global value of the density parameter, $\Omega_0$, in 
equation~(\ref{sigmapeak}), with a local $\Omega_{\rm eff}$.  
Once this adjustment has been made, linear theory provides a good 
description of the distribution of halo speeds.  
In addition, linear theory provides a good description of the directions 
of halo motions.  That is to say, once $\Omega_0\to\Omega_{\rm eff}$, 
an approach based on the Zeldovich approximation should be reasonably 
accurate at describing halo motions.  The same is not true, of course, 
for the motions of dark matter particles; particles receive substantial 
nonlinear kicks as they become incorporated into clusters.  These kicks 
are approximately randomly distributed in direction.  

\subsection{Comparison with simulations:  Virial motions}
We have assumed throughout that the virial term depends on mass 
more strongly than it does on local density.  For completeness, we 
think it useful to show that this is accurate.  

\begin{figure}
\centering
\mbox{\psfig{figure=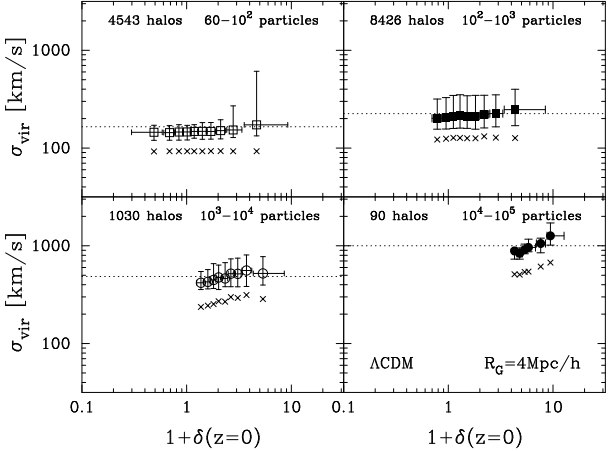,height=6.cm,bbllx=82pt,bblly=62pt,bburx=608pt,bbury=452pt}}
\caption{Dependence of virial motions on halo mass and local density in 
the $\Lambda$CDM simulations.  Symbols in the top left panel show 
the lowest mass haloes, symbols in the bottom right panel show the most 
massive haloes.  Dotted curve shows equation~(\ref{sigmavir}) computed 
for the median halo mass in the panel, and using $\Omega_0$, the global, 
not the local, value of the density.  Crosses (offset downwards for clarity) 
show the result of inserting the median mass in each bin into 
equation~(\ref{sigmavir}) to compute $\sigma_{\rm vir}$.}
\label{virdelta}
\end{figure}

The symbols with error bars in the four panels in Fig.~\ref{virdelta} 
show $\sigma_{\rm vir}$ as a function of local density for $\Lambda$CDM 
haloes in four different mass ranges; least massive haloes are top 
left, and most massive haloes are bottom right.  The mass cuts were 
chosen to be the same as in Fig.~\ref{assumel}, so the symbols in 
this figure are the same as in that one.  And as before, the error 
bars show the range in each bin within which 90\% of the haloes are.  
The density was defined by 
smoothing the particle distribution on a grid with a Gaussian filter 
of width 4Mpc/$h$.  The dotted curves show the value of $\sigma_{\rm vir}$ 
predicted by the virial theorem (our equation~\ref{sigmavir}) for the 
associated mass range.  The dotted curves were computed using the 
global value of the density parameter $\Omega_0$, not the local one.  
They provide a reasonable fit to the simulations.  
For massive haloes (bottom panels), $\sigma_{\rm vir}$ appears to be 
higher in denser regions.  
This is almost entirely a consequence of the fact that the densest 
cells contain the more massive haloes in the mass bin.  To illustrate 
this, we have computed the median halo mass in each density bin, 
using only the haloes in the requisite mass range for each panel, 
and used this mass to compute $\sigma_{\rm vir}$.  These values are 
shown as the crosses in each panel---they have been offset 
downwards by a factor of 0.25 (i.e., the plot shows the virial 
velocity divided by $10^{0.25}$) for clarity.  The crosses show the 
same trend as the main symbols, demonstrating that the virial term 
really does depend on mass, and not on local density.  

\section{Discussion}\label{discuss}

We described a model in which galaxy velocities were written as the 
sum of two terms, one of which was effectively described using linear 
theory, and the other which was inherently nonlinear.  We showed that 
this split was both useful and accurate for clustering which develops 
from Gaussian initial conditions, although our model should also work 
for more general initial conditions.  For Gaussian initial conditions, 
a generic prediction of our model is a Gaussian core, with exponential 
wings for the distribution of galaxy peculiar velocities.  
The core mainly reflects the fact that most dark matter haloes 
move with the same speed, and this speed is less than the virial motions 
within clusters (both these are consequences of the Gaussian initial 
conditions); the wings arise primarily from the large nonlinear virial 
velocities within massive clusters.  
We compared our model with one dimensional peculiar velocity 
distribution functions $f(v)\,{\rm d}v$ extracted from dissipationless 
N-body simulations, where galaxies were formed and evolved with 
semi-analytic techniques. Our model, which explicitly considers whether 
dark matter particles, galaxies or galaxy clusters trace the velocity 
field, accounts for the shape of $f(v)\,{\rm d}v$ extremely well.

We are not the first to have considered the shape of the galaxy 
distribution function statistic.  Saslaw et al. (1990) present a 
formula for $f(v)\,{\rm d}v$ which is accurate for N-body simulations 
of clustering from Poisson initial conditions.  
The virtue of our approach is that it shows clearly how the 
distribution should depend on trace particle type.  
Raychaudhury \& Saslaw (1996) present a measurement of $f(v)\,{\rm d}v$ 
using a sample of spirals compiled by Mathewson, Ford \& Buchhorn (1992).  
It would be interesting to do the same with the more recent SFI and ENEAR 
data sets of spirals and ellipticals.  Since our models depend explicitly 
on cosmology, it may be that these data sets are able to place 
interesting constraints on the density parameter.  

Section~\ref{evolve} argued that halo speeds evolve faster in denser 
regions.  The fact that massive haloes populate denser regions has
consequences for estimates of $\Omega_0$, studies of how and why the 
local value of the cosmic Mach number may differ from the global one, 
and the detection of the kinematic Sunyaev-Zeldovich effect.  

Suppose one were to measure the peculiar velocities of galaxies.  
Since measuring distances is difficult, the velocity of any one galaxy 
is often ill-determined.  To increase signal to noise, one often `groups' 
galaxies, before using their `grouped' velocities.  If these groups 
correspond approximately to dark matter haloes, then this is useful, 
because we showed that linear theory describes the directions of halo 
motions rather well.  However, one often then weights all subsequent 
analyses by the signal-to-noise ratio.  In effect, this means that the 
signal-to-noise weighted grouped velocities are dominated by the motions 
of massive haloes.  Since the motions of massive haloes evolve more 
quickly than average, they have a higher effective density parameter than 
the global value.  It is interesting, therefore, that peculiar velocity 
based estimates of $\Omega_0$ are often higher than those indicated by 
almost all other observations (e.g., Smith et al. 1999; 
Branchini et al. 2000; Zaroubi et al. 2000; but see Borgani et al. 2000).  

The cosmic Mach number is the ratio of the large scale bulk velocity 
to the smaller scale velocity dispersion (Ostriker \& Suto 1990).
Recent work (van de Weygaert \& Hoffman 1999) suggests that local 
measurements of this Mach number may differ from the global one.  
Because our model is also phrased in terms of bulk flows and dispersions, 
it can be used to address this issue analytically.  For example, it 
allows one to study how the distribution of the Mach number statistic 
depends on local density.  
 
An obvious consequence of our results is that large overdense 
regions, namely superclusters, are ideal places to look for the 
kinematic Sunyaev-Zeldovich effect:  in these regions, low mass haloes, 
with relatively low optical depth, can have substantial peculiar 
velocities; they thus can yield detectable kinematic Sunyaev-Zeldovich 
fluctuations in the Cosmic Microwave Background radiation 
(Diaferio, Sunyaev and Nusser 2000).  In this context, our model for 
the distribution of halo speeds, and its dependence on density, is 
then related to the distribution of temperature fluctuations one might 
measure.  

Our results suggest that whereas nonlinear effects can be thought 
of as adding Gaussian noise to the initial linear theory velocity, 
this noise is higher for particles in massive haloes.  Therefore, the 
noise is spatially dependent---adding or subtracting a global thermal 
noise before comparing observations with theory may lead to 
inconsistencies.  Should one choose to use a global thermal noise 
anyway (e.g. Freudling et al. 1999), then the `temperature' associated 
with ellipticals (e.g. the ENEAR catalog), which occur more in clusters, 
will be larger than that for spirals (e.g., the SFI catalog) which occur 
more in the field.  

Our derivation of the single particle velocity distribution function 
is the first step in a larger program.  Baker, Davis \& Lin (2000), 
following Davis, Miller \& White (1997), consider a statistic which is 
essentially a smoothed version of our $f(v)$ distribution.  
The operation of smoothing is essentially one of adding several, possibly 
correlated, velocities.  So, to compute the smoothed statistic from our 
model requires knowledge of how the halo velocity correlation function 
depends on halo mass.  We are in the process of extending our model to 
include the effects of smoothing.  

Finally, our results suggest that, once account has been taken for 
the effect of the local density on the evolution of velocities, 
linear theory (or the Zeldovich approximation) should provide a 
reasonably good estimate of the motions of haloes.  This simplifies 
models of redshift-space distortions considerably (work in progress).  
Catelan et al. (1998) describe a dark-matter halo based model of the 
density field in which motions are approximated using the Zeldovich 
approximation.  Our results suggest that if their approach were to be 
extended to describe redshift-space distortions, then it should be quite 
accurate.  

\section*{ACKNOWLEDGMENTS}
Thanks to Enzo Branchini, Mike Hudson, Mariangela Bernardi, 
Michael Blanton, J\"org Colberg, Lam Hui, Rom\'an Scoccimarro, 
Rien van de Weygaert, Saleem Zaroubi and Idit Zehavi for helping 
shape our thoughts on these subjects.  
The N-body simulations, halo and galaxy catalogues used in this paper
are publically available at {\tt http://www.mpa-garching.mpg.de/NumCos}.
The simulations were carried out at the Computer Center of the 
Max-Planck Society in  Garching and at the EPCC in Edinburgh, as part 
of the Virgo Consortium project.  In addition, we would like to thank 
the Max-Planck Institut f\"ur Astrophysik where this project began, and 
where some of the computing for this work was done, and Simon White 
for prompting us to think about the question of isotropy.  
RKS is supported by the DOE and NASA grant NAG 5-7092 at Fermilab.  
He thanks the Department of Physics as well as the Observatory at 
Torino for their warm hospitality in May 2000.

\appendix
\section{The density in cells containing haloes}
Massive haloes populate denser cells:  
this Appendix provides a simple estimate of how the mean 
density in cells which contain at least one halo of mass $m$ 
increases as $m$ increases.  

Let $p(\delta)\,{\rm d}\delta$ denote the probability that a cell of 
size $V$ has overdensity in the range ${\rm d}\delta$ about $\delta$, 
and let $p(j|\delta)$ denote the probability that such a region in 
which the overdensity is $\delta$ contains $j$ haloes of mass $m$.
The mean overdensity within cells of size $V$ which contain at least 
one halo of mass $m$ is 
\begin{displaymath}
\Bigl\langle\delta|{\rm at\ least\ one\ halo}\Bigr\rangle = 
{\int {\rm d}\delta\ \delta\,p(\delta)\ [1-p(j=0|\delta)] \over 
\int {\rm d}\delta\ p(\delta)\ [1-p(j=0|\delta)]},
\end{displaymath}
where the denominator is the fraction of cells that contain at least
one halo of mass $m$.  

If $p(j|\delta)$ were Poisson, then the probability that $j=0$ would 
be ${\rm e}^{-n(m|\delta)V}$.  Sheth \& Lemson (1999) show that this 
is a reasonable assumption on large scales.  Since there are many more 
low-mass haloes than massive ones, this exponential term tends to zero 
for small masses, and so the integrals above reduce to 
$\langle\delta\rangle = 0$.  This says that the average overdensity 
in cells containing at least one less massive halo is zero.  
Massive haloes are slightly more complicated.  To see how the 
integrals scale, it is useful to consider the large scale limit.  
In this limit, the linear bias model is reasonably accurate:  
 $n(m|\delta)\approx n(m)\,[1 + b(m)\delta]$, 
where $b(m)$, the linear bias factor, increases as the mass $m$ of 
the haloes of interest increases.  To simplify things even further, 
suppose that $n(m|\delta)V\ll 1$ (although this assumption is 
inconsistent with the assumptions of linear bias and Poisson statistics, 
it helps illustrate our argument more clearly).  
Then $1-p(j=0|\delta)\approx n(m)V\,[1 + b(m)\delta]$ and the 
integrals above become 
\begin{displaymath}
\Bigl\langle\delta|{\rm at\ least\ one\ halo}\Bigr\rangle = 
{\bigl\langle \delta\,n(m)V\,[1 + b(m)\delta] \bigr\rangle \over  
\bigl\langle n(m)V\,[1 + b(m)\delta] \bigr\rangle} = 
b(m)\,\bigl\langle\delta^2\bigr\rangle, 
\end{displaymath}
where we have used the fact that $\langle\delta\rangle \equiv 0$.  
Since $b(m)$ increases as $m$ increases, this says that at fixed $V$, 
the mean density of cells in which there is at least one 
halo of mass $m$ increases as $m$ increases.  
This is in qualitative agreement with Fig.~\ref{v0d0r4} which shows 
that the median density of cells which contain at least one halo of 
mass $m$ increases as $m$ increases.

\end{document}